\documentclass[article]{aa}  

\usepackage{newtxtext,newtxmath}

\usepackage[T1]{fontenc}
\usepackage{ae,aecompl}
\usepackage{comment}
\usepackage{aas_macros}
\usepackage[hidelinks]{hyperref}
\usepackage{subcaption}

\usepackage{graphicx}	
\usepackage[export]{adjustbox}

\usepackage{xcolor}
\usepackage{multicol}
\usepackage{multirow}

\def\gsim{\;\raise0.3ex\hbox{$>$\kern-0.75em\raise-1.1ex\hbox{$\sim$}}\;}
\def\Msun{M_\odot}

\begin{document}

\title{X-ray flux -- mass relation for $z\gtrsim 0.7$ galaxy clusters}

\author{
N.~Lyskova \inst{1}
\and
E.~Churazov \inst{2,1} 
\and
I.~I.~Khabibullin \inst{3,1,2} 
\and
R.~Sunyaev \inst{1, 2}
\and
M. Gilfanov \inst{1,2}
\and
S. Sazonov \inst{1}
}

\institute{
Space Research Institute (IKI), Profsoyuznaya 84/32, Moscow 117997, Russia
\and
Max Planck Institute for Astrophysics, Karl-Schwarzschild-Str. 1, D-85741 Garching, Germany 
\and 
Universitäts-Sternwarte, Fakultät für Physik, Ludwig-Maximilians-Universität München, Scheinerstr.1, 81679 München, Germany
}

\abstract{We use a subsample of co-detections of the ACT and MaDCoWS cluster catalogs to verify the predicted relation between the observed  X-ray flux $F_{\rm X}$ in the 0.5-2~keV band and the cluster mass $M_{\rm 500c}$ for halos at $z>0.6-0.7$. We modify this relation by introducing a correction coefficient $\eta$, which is supposed to encapsulate factors associated with a particular method of flux estimation, the sample selection function, the definition of the cluster mass, etc.  
We show that the X-ray flux, being the most basic X-ray observable, serves as a convenient and low-cost mass indicator for distant galaxy clusters with photometric or even missing redshifts (by setting $z=1$) as long as it is known that $z\gtrsim 0.6-0.7$. The correction coefficient  $\eta$ is $\approx 0.8$ if $M^{\rm UPP}_{\rm 500c}$ from the ACT-DR5 catalog  are used as cluster masses and $\eta\approx 1.1$ if weak-lensing-calibrated masses $M^{\rm Cal}_{\rm 500c}$ are used instead.  }

\titlerunning{Fx-M500c}

    
\maketitle

\section{Introduction}
\label{s:intro}

In the current paradigm of structure formation, galaxy clusters are formed through the gravitational collapse of the largest (and rarest) density peaks in the primordial density field. Thus, the abundance of clusters (as a function of redshift) carries the imprint of growth of structure over cosmological time and, as a consequence, serves as a sensitive probe of the underlying cosmological parameters \citep[for reviews see, e.g.,][]{2011ARA&A..49..409A, 2012ARA&A..50..353K,2019SSRv..215...25P}. Over the past several decades, multi-wavelength observations of galaxy clusters have been a major focus of research. Clusters can be efficiently detected in the optical, X-ray, and the Sunyaev-Zeldovich (SZ; \citealt{1972CoASP...4..173S}) effect observations. Numerous galaxy cluster surveys deliver (and will be delivering) large samples of cluster candidates based on different observational proxies, e.g., surveys from the Dark Energy Spectroscopic Instrument (\citealt{2016arXiv161100036D}), the eROSITA telescope on board the SRG observatory \citep{2021A&A...656A.132S,2021A&A...647A...1P}, and Atacama Cosmology Telescope (ACT; \citealt{2011ApJS..194...41S}), to name a few. The main scientific driver for these surveys is to constrain cosmological parameters by measuring the cluster mass function, for which reliable mass measurements for clusters are necessary. Numerical simulations can predict the abundance of massive halos \citep[e.g.,][]{2001MNRAS.323....1S,2001MNRAS.321..372J,2008ApJ...688..709T,2006ApJ...646..881W} taking into account background cosmology and additional effects, e.g., active galactic nuclei (AGN)  feedback. While different cluster properties can be predicted, most often the mass function is used as a compact proxy to cosmological parameters, e.g., the amplitude of perturbations at scales $\sim 10\,{\rm Mpc}$ or the growth factor as a function of $z$.  For a fair comparison, one would like to derive the cluster mass function from observations. However, the mass cannot be measured directly. Instead, we often rely on scaling relations between the cluster mass and observable properties, such as optical richness, velocity dispersion, integrated Compton-y parameter, X-ray luminosity, gas temperature, gas mass, etc. \citep[e.g.,][]{2013SSRv..177..247G}. Ideally, one would like to have an observable (or a set of observables) that are (1) easy to measure, (2) accessible for large samples of clusters (i.e., no additional observations are required), (3) tightly related to total cluster mass, and (4) having low scatter. Another useful observable would be an observable that is readily available and can be used for a pre-selection of cluster samples for deeper studies. This is the main focus of the present study, with the observed X-ray flux taking the role of such an observable.   

Cluster masses could be determined from, for example, weak lensing (WL) analysis. Currently, such an analysis is considered to be one of the least biased methods to measure the total masses of massive clusters \cite[][]{2016MNRAS.463.3582M, 2020A&ARv..28....7U, 2025PJAB..101..129O}. However, there are still inherent systematic errors (for instance, triaxiality, projection effects, miscentering, etc., e.g., \citealt{2013MNRAS.429..661M}) and, from the observational point of view, WL requires large-area observations of background galaxies.    

Most commonly, cluster masses are inferred from X-ray and SZ observations. The total SZ flux is expected to correlate tightly with the total cluster mass \cite[e.g.,][among many others]{2006ApJ...650..538N, 2010A&A...517A..92A, 2013A&A...550A.129P}. X-ray observations of hot intracluster gas provide information about gas temperature and density profiles, enabling the determination of individual cluster masses under the assumption of hydrostatic equilibrium. However, such detailed X-ray observations are not feasible for large samples of clusters, and for shallow surveys, global cluster properties such as X-ray luminosity or temperature, or gas mass can be used to estimate the total mass. Among X-ray mass proxies, $Y_{\rm X} = M_{\rm gas}T$, which is a product of gas mass and X-ray spectroscopic temperature, is believed to have the lowest scatter \citep{2006ApJ...650..128K}. Although, when data quality does not permit to determine temperature of a cluster, we have to rely on maybe less accurate but easier to measure observables such as X-ray luminosity \citep[e.g.,][]{2009A&A...498..361P,2018MNRAS.473.3072M,2022A&A...665A..24P} or flux \citep{2015MNRAS.450.1984C}, or the average energy of the X-ray spectrum - an alternative to X-ray temperature for low-redshift clusters suggested in \cite{2024arXiv240812026K}. Here, we focus on a sample of high-redshift ($z>0.7$) galaxy clusters \citep{2021A&A...653A.135O} with SZ-based mass measurements and take advantage of the all-sky SRG/eROSITA survey data to verify and calibrate the scaling relation between the cluster mass and the X-ray flux. The latter is the most direct X-ray observable,  which can be used for mass estimates even if redshift is not known (for distant objects). 

The paper is organized as follows: Section~\ref{sec:data} provides a description of the data and catalogues we use. In Section~\ref{sec:3}, we present the baseline relation between the cluster mass and the X-ray flux and how eROSITA X-ray fluxes are measured. Next, we calibrate the $F_{\rm X}-M_{\rm 500c}$ relation in Section~\ref{sec:4} and summarize the obtained results in Conclusions.

This work relies on X-ray scaling relations obtained/presented in \cite{2009ApJ...692.1033V} and \cite{2015MNRAS.450.1984C}. For consistency, we adopt the same set of cosmological parameters as in \cite{2015MNRAS.450.1984C}: $\Omega_M=0.27$, $\Omega_\Lambda=0.73$, $h=0.7$.
Throughout the paper, a cluster mass refers to $M_{\rm 500c}$, which is defined within the radius $R_{\rm 500c}$ enclosing the average mass density of 500 times the critical density at a cluster redshift.

\section{Data sets}
\label{sec:data}

\subsection{eROSITA data preparation}
\label{sec:eROdata}
We use the data of the eROSITA telescope \citep{2021A&A...647A...1P} on board the  \textit{SRG} mission  \citep{2021A&A...656A.132S} accumulated over four consecutive all-sky surveys.
Initial reduction and processing of the data were performed at IKI using standard routines of the \texttt{eSASS} software \citep{2018SPIE10699E..5GB,2021A&A...647A...1P} and proprietary software developed in the RU eROSITA consortium, while the imaging and spectral analysis were carried out with the background modeling, vignetting, point spread function (PSF) and spectral response function calibrations built upon the standard ones via slight modifications motivated by results of calibration and performance verification observations \citep[e.g.,][]{2021A&A...651A..41C,2023MNRAS.521.5536K}.

\subsection{SZ cluster catalogues}
\label{sec:SZdata}
Recently, a large catalog of Sunyaev-Zel'dovich (SZ) selected galaxy clusters surveyed by the Atacama Cosmology Telescope (ACT) has been published by \cite{2021ApJS..253....3H}. The ACT DR5 catalog provides cluster mass estimates $M_{\rm 500c}$ derived from an SZ signal for a large sample of clusters. At the redshift range of our interest $z>0.7$, there are 364 clusters with the Galactic longitude 0$^{\circ}$ < $l$ < 180$^{\circ}$
and with available SZ-based masses. As a mass proxy we use values $M_{\rm 500c}^{\rm UPP}$ (the column \texttt{M500c}  in the ACT DR5 catalog) which were obtained by assuming the universal pressure profile and the scaling relation between an SZ signal and a cluster mass from  \cite{2010A&A...517A..92A}.
\cite{2021ApJS..253....3H}  report also $M_{\rm 500c}^{\rm Cal}$ (the column \texttt{M500cCal}) masses rescaled using the richness-based weak-lensing mass calibration, where $M_{\rm 500c}^{\rm Cal}$ =  $M_{\rm 500c}^{\rm UPP}/0.71$  \citep{2021ApJS..253....3H}. 

For the calibration of the relation between an X-ray flux and a cluster mass, we make use of a catalog from \cite{2021A&A...653A.135O}, which consists of clusters detected both in the  Massive and Distant Clusters of WISE Survey (MaDCoWS; \citealt{2019ApJS..240...33G}) and ACT data. Keeping objects
with the galactic longitude 0$^{\circ}$ < $l$ < 180$^{\circ}$ and $z \ge 0.7$ yields 36 clusters. 
Cluster masses obtained in \cite{2021A&A...653A.135O} are essentially identical  with $M_{\rm 500c}^{\rm UPP}$ values from \cite{2021ApJS..253....3H}. 

\section{X-ray flux as the mass proxy}
\label{sec:3}
\subsection{$F_{\rm X}$-$M_{\rm 500c}$ relation}
\label{sec:FxM}

\begin{figure*}
    \centering
    \includegraphics[angle=0,clip=true,width=0.65\columnwidth]{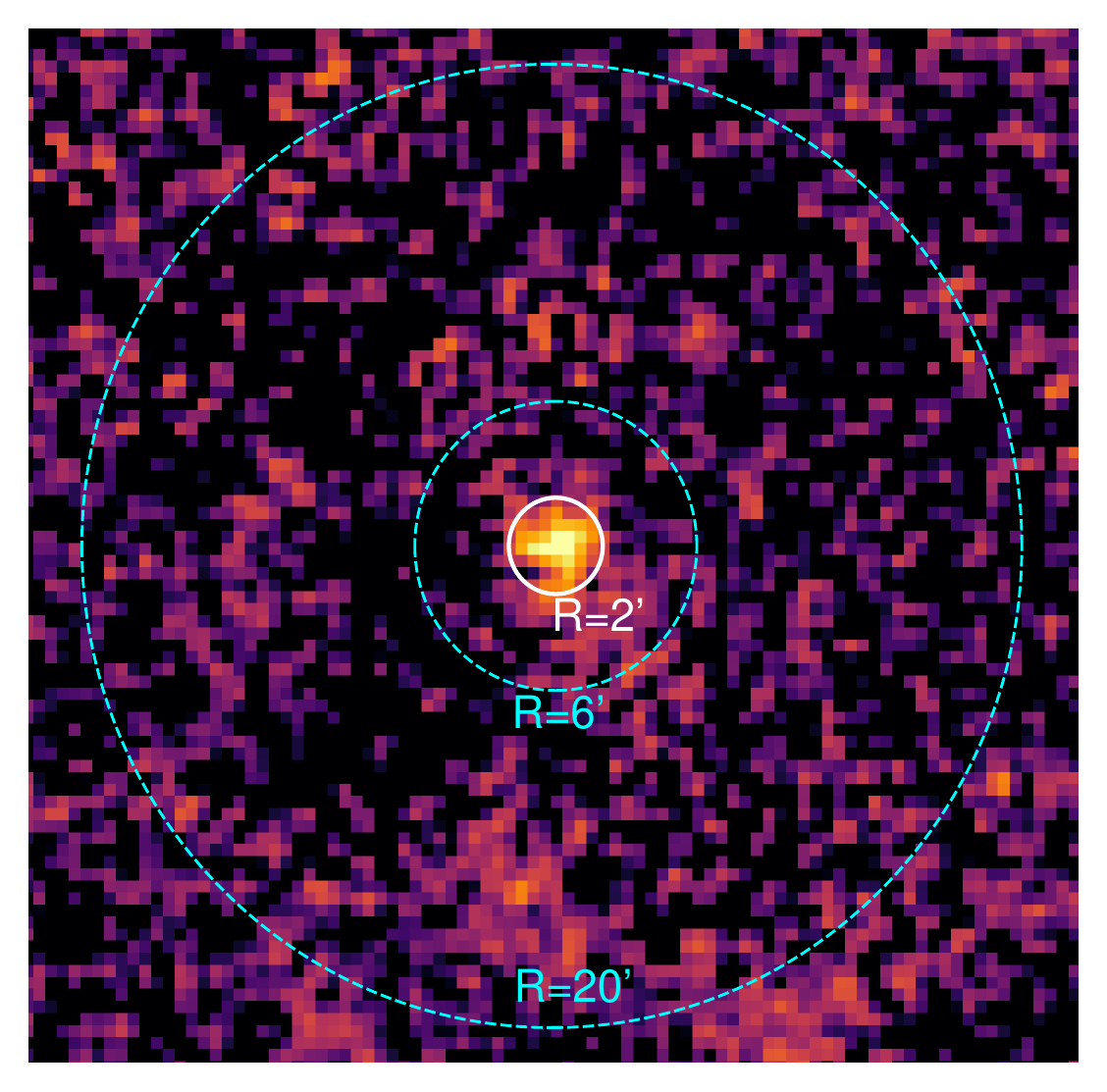}
    \includegraphics[angle=0,clip=true,width=0.65\columnwidth]{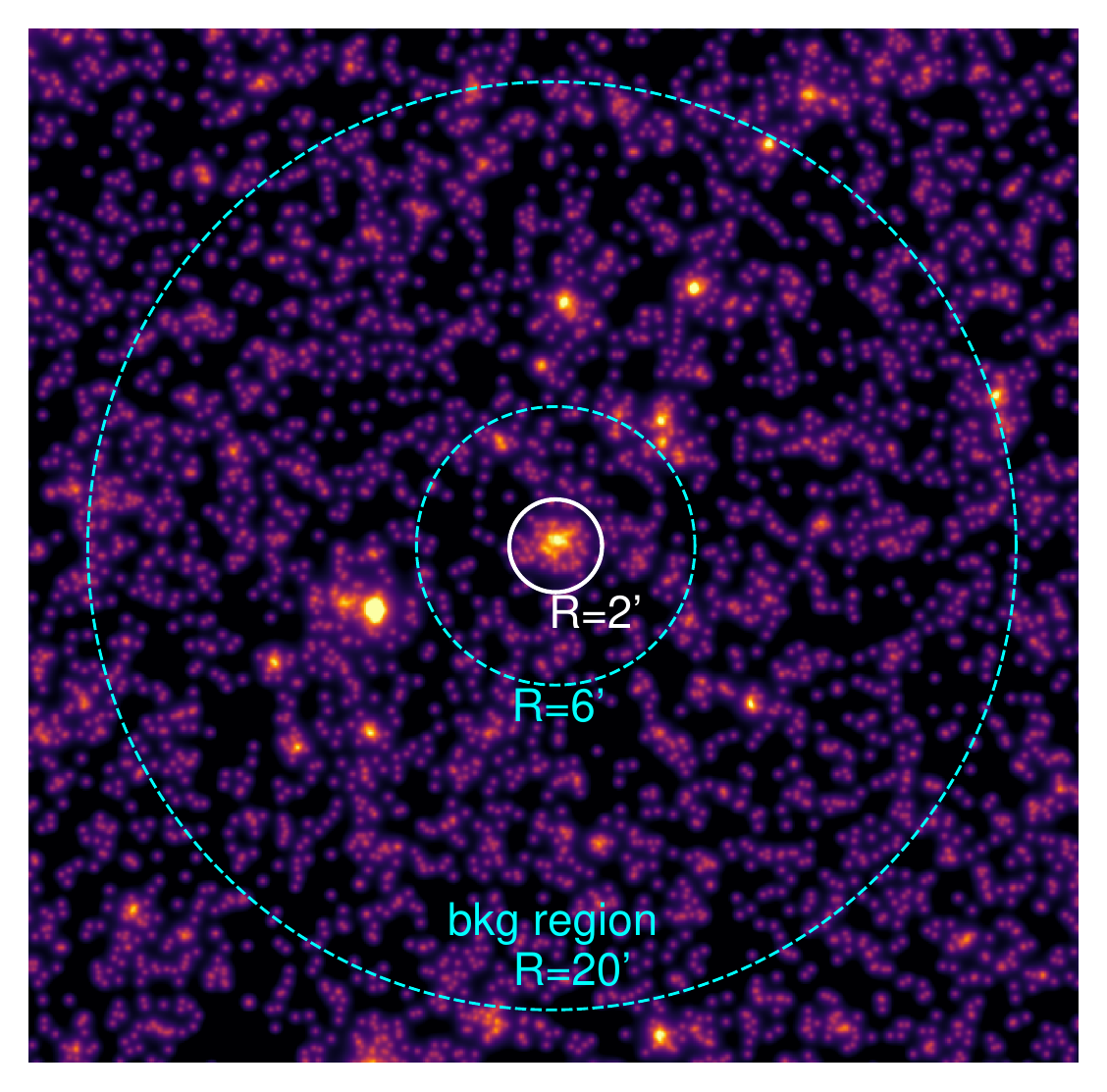}
    \includegraphics[angle=0,clip=true,width=0.65\columnwidth]{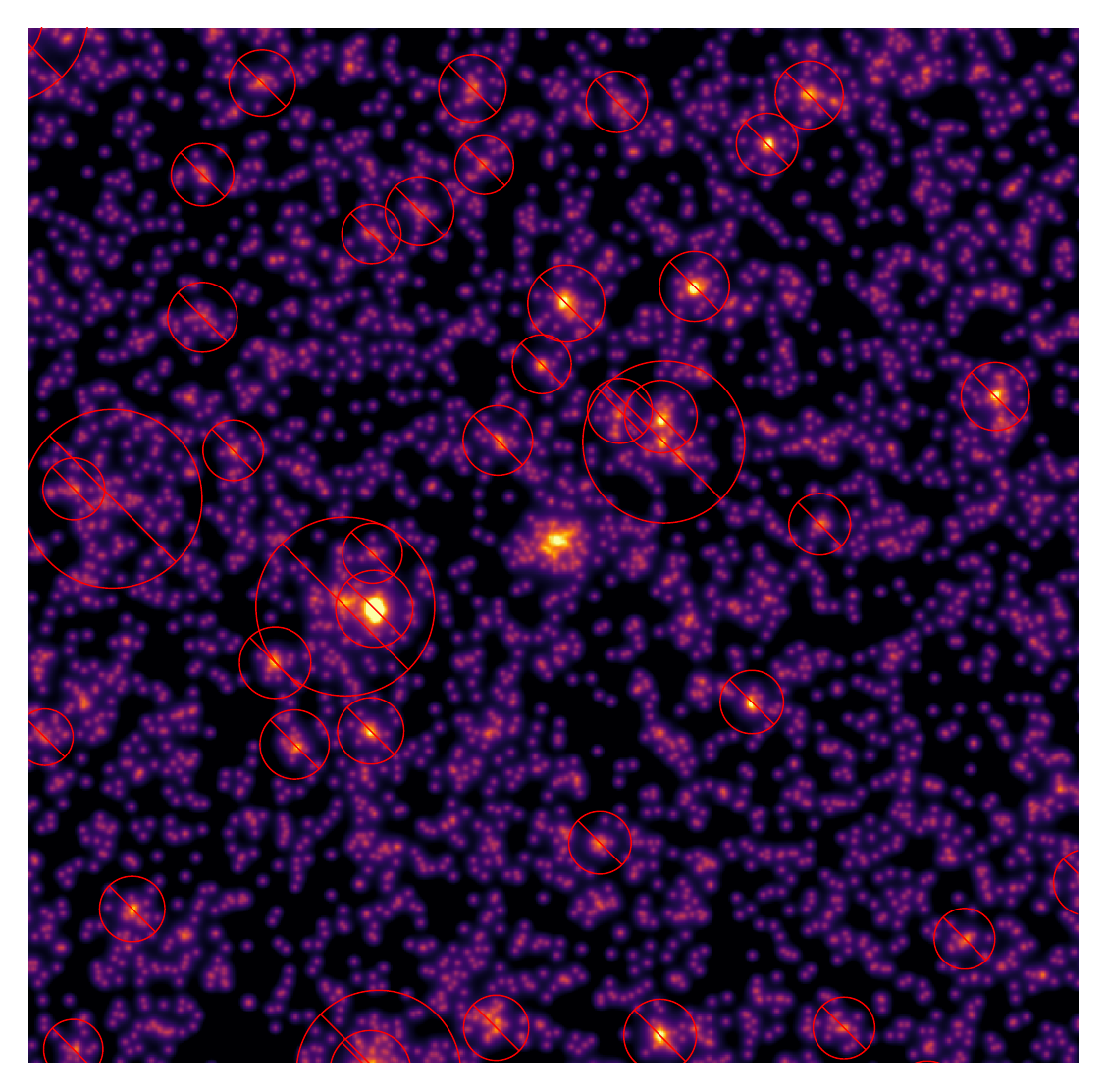}    
    \caption{SZ and X-ray views on the galaxy cluster ACT-CL J0105.8-1839. Left panel: the Planck + ACT  Compton-y map (\citealt{2024PhRvD.109f3530C}). Middle panel: the eROSITA 0.4--2.3 keV image. We use a ring with $6'<R<20'$ to estimate the background, which is then subtracted from the source region marked as a white circle. The right panel illustrates the source removal procedure.}
    \label{fig:X_SZ_example}
\end{figure*}

\cite{2015MNRAS.450.1984C} showed that for $z > 0.6$  there is a relatively tight correlation between the observed X-ray flux in the 0.5-2.0 keV energy range and the cluster mass $M_{\rm 500c}$:
\begin{equation}
M_{\rm 500c} =    1.2\times 10^{14} M_\odot\; \left ( \frac{F_{\rm X}}{10^{-14}\,{\rm erg\,s^{-1}cm^{-2}}}\right )^{0.57}.
\label{eq:CVS2015}
\end{equation}
The above expression is based on the X-ray scaling relations in \cite{2009ApJ...692.1033V}, with its tightness resulting from the evolution of cluster properties with redshift.  We use it as a baseline to derive a simple recipe for estimating galaxy cluster masses from eROSITA data. 
We patch the relation~(\ref{eq:CVS2015}) with a weak dependence on the redshift and add a correction coefficient $\eta$ which is supposed to encapsulate factors associated with, for example,  a particular method of flux estimation, the sample selection function, and the definition of the mass.
The modified $F_{\rm X}$ - $M_{\rm 500c}$ relation then reads as:
\begin{equation}
M_{\rm 500c} =    1.2\times 10^{14} M_\odot\; \eta\;\left ( \frac{F_{\rm X}}{10^{-14}\,{\rm erg\,s^{-1}cm^{-2}}}\right )^{0.57}\;z^{0.5}.
\label{eq:xmass}
\end{equation}
The scaling relations of \cite{2009ApJ...692.1033V} were derived from the analysis of X-ray selected clusters with masses $M_{\rm 500c} \gtrsim 1.5\times 10^{14}\, M_{\odot}$ at $z=$ 0.02 - 0.9. Here, we consider a similar mass range and assume that the scaling relations hold for $ 0.7 < z < 2$. Then, the factor  $z^{0.5}$ might improve the tightness of the relation~(\ref{eq:CVS2015}) (see Fig. 4 in \citealt{2015MNRAS.450.1984C} and Appendix~\ref{sec:z05}), unless it is subdominant to the observational or intrinsic scatter in clusters’ properties. 
This factor can be omitted for crude mass estimates, which is the main purpose of this procedure.

\subsection{X-ray flux estimation}
\label{sec:Fx}

Even the most massive clusters at redshift $z\sim 1$ are expected to be relatively faint X-ray sources, with the flux $\rm\lesssim few\times 10^{-13} \,{ \rm erg\,s^{-1}\,cm^{-2}}$ in 0.5--2 keV \citep[e.g.,][their Fig.~3]{2015MNRAS.450.1984C}, so, with a shallow exposure of the eROSITA all-sky survey at most $\sim100$ counts is expected to be detected from them. Although this is well above the commonly accepted detection limits due to pure statistical (Poisson) and background fluctuations, the robust characterization of faint extended sources is more challenging. Moreover, the diffuse X-ray emission from galaxy clusters can be contaminated by an (unresolved) AGN, which makes flux measurements via simple fitting strongly dependent on the spatial model assumptions. Instead, we measure X-ray fluxes from a fixed aperture. Such an approach is model-independent, simple in use, and fast, but not without its own flaws, which will be discussed below in Appendix~\ref{sec:bias}.

Assuming the Tinker mass function \citep{2008ApJ...688..709T}, we estimated that the top most massive 100 clusters at $z \geq 0.7$ have masses $M_{\rm 500c}\sim 6.5\times 10^{14}$ M$_{\odot}$. The angular sizes of the $R_{\rm 500c}$ regions, where the bulk of the X-ray emission is expected to be concentrated, $\sim 2.4'$. For comparison, for most massive 100 clusters at $z \geq1$,  $M_{\rm 500c} \sim 4.5\times 10^{14}$ M$_{\odot}$ and  $R_{\rm 500c} \sim 1.6'$. Below, we estimate the X-ray flux from galaxy clusters using a circular aperture with fixed size: its radius is set to $r=2'$.

First, we measure the 0.4--2.3 keV count rate within a circle with radius $r=2'$ centered at the cluster position. The local X-ray background signal is estimated in a ring with  $6' < r < 20'$  (see Fig.~\ref{fig:X_SZ_example} for an illustration). To reduce scatter in the flux estimates associated with the Cosmic X-ray Background (CXB), we detect and mask bright sources using flux cuts discussed in the following sections.  The detection and characterization of sources are performed in the same way as in \cite{2021A&A...651A..41C,2023MNRAS.521.5536K},
where special care was taken to separate the extended diffuse emission from the point or mildly extended sources. Next, the extracted count rates within the source apertures are corrected for the background contributions and converted to the 0.5--2 keV flux using the constant factor, which is suitable for extragalactic sources with hot thermal spectrum and small absorbing column density \citep[e.g.,][]{2023MNRAS.525..898L}. The resulting flux is further corrected for the excess in the resolved CXB fraction in the background region compared to the source aperture (see Appendix~\ref{sec:bias}). Such a correction is needed if there is a non-negligible bias arising from different flux cuts applied to source and background regions.

\section{Calibration of the $F_{\rm X}$-$M_{\rm 500c}$ relation}
\label{sec:4}
\subsection{ACT/MaDCoWS sample}
\label{sec:calib}

\begin{figure}
    \centering
    \includegraphics[angle=0,clip=true,width=\columnwidth]{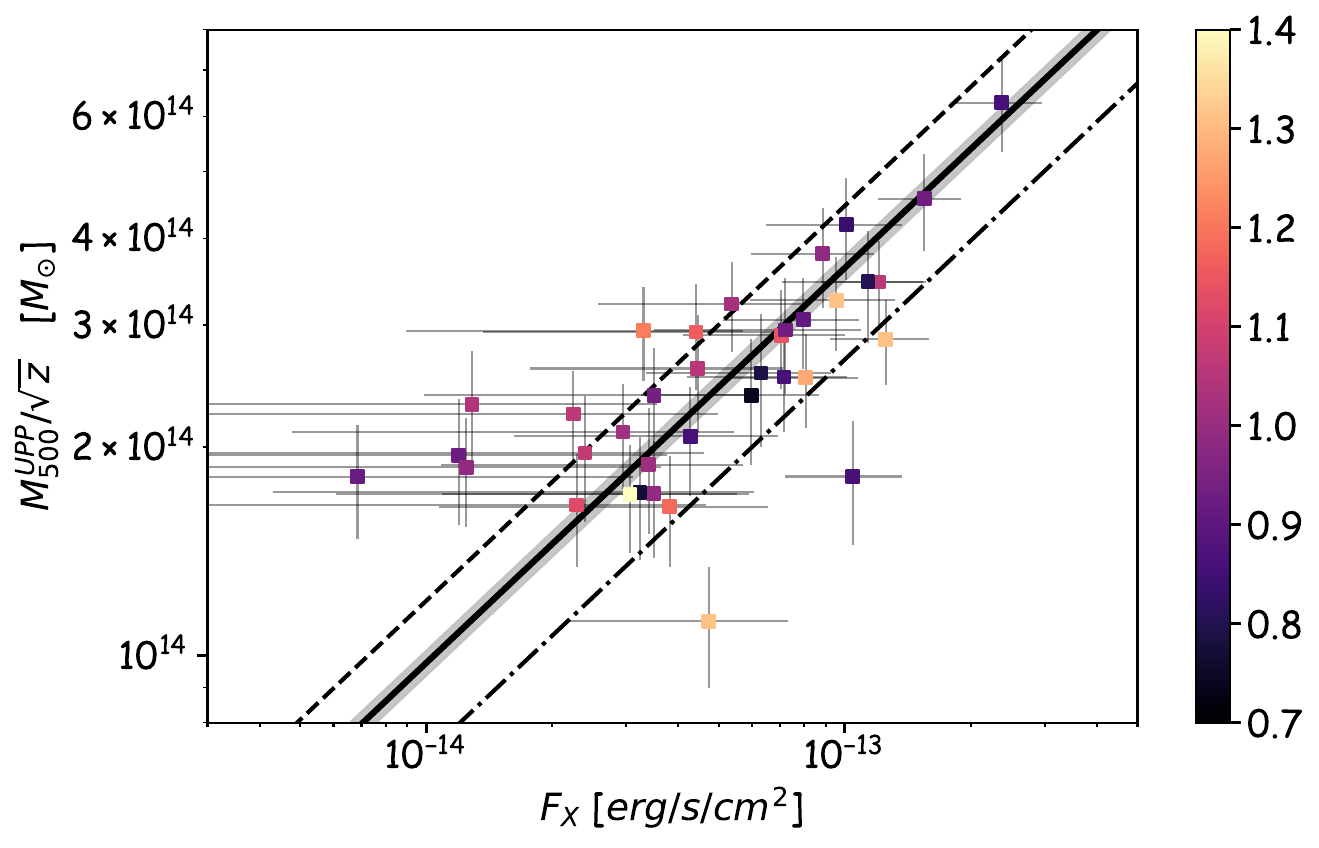}       
    \caption{ACT cluster masses $M^{\rm UPP}_{\rm 500c}$ vs the eROSITA fluxes in the 0.5--2.0 keV energy band for the subsample of ACT/MaDCoWS co-detections from \citealt{2021A&A...653A.135O}. X-ray fluxes have been corrected for the expected bias arising from the different resolved fractions of compact sources in the cluster and background regions (see Appendix~\ref{sec:bias}).
    The best-fit relation~(\ref{eq:xmass}) with $\eta = 0.8\pm 0.03$  is shown with a black line and grey shaded area. The dashed and dash-dotted lines illustrate $\eta=1$ and 0.6, respectively. Cluster redshifts are shown with color.}  
    \label{fig:overview}
\end{figure}

\begin{figure}
     \includegraphics[angle=0,clip=true,width=\columnwidth]{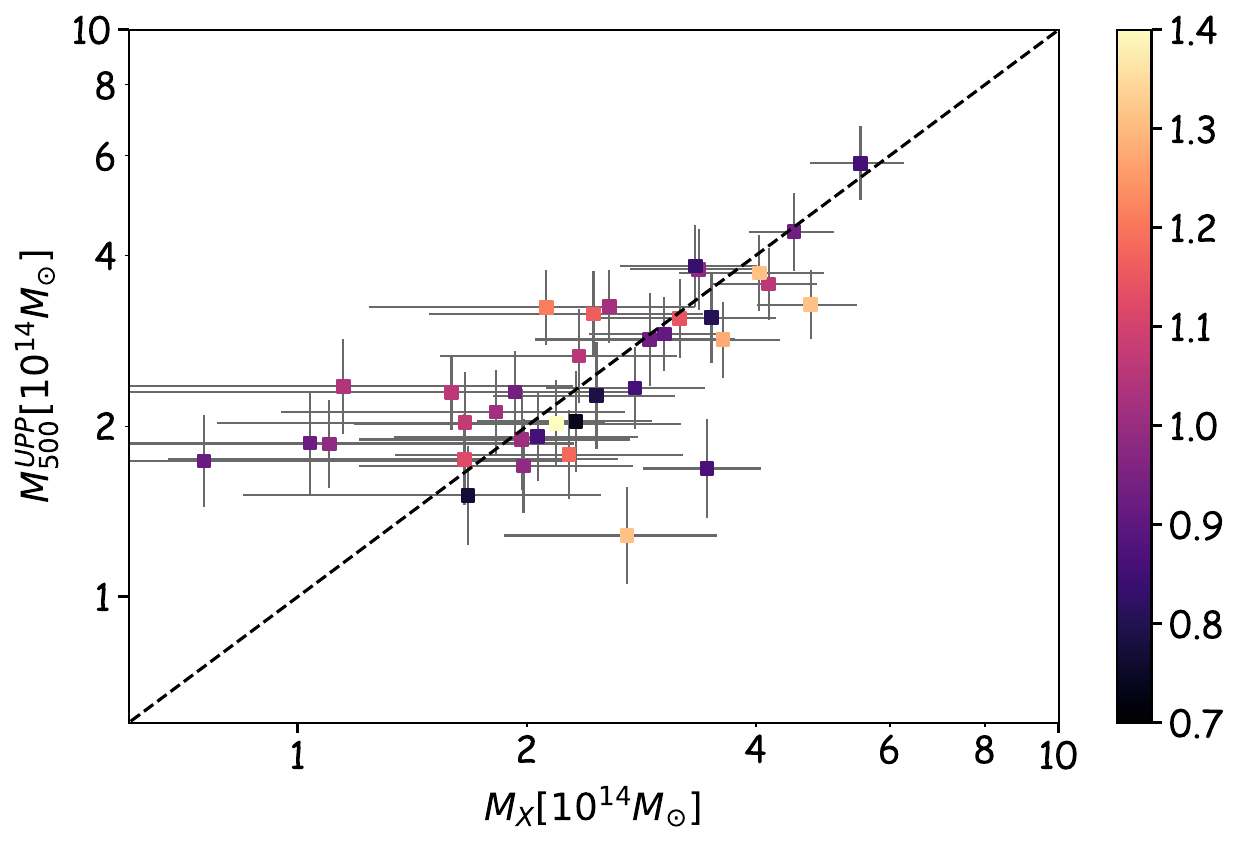}
    \caption{ACT masses $M^{\rm UPP}_{\rm 500c}$ vs cluster masses estimated from the calibrated $F_{\rm X}$-$M_{\rm 500c}$ relation for the \cite{2021A&A...653A.135O} subsample.  Cluster redshifts are shown with color.  }
    \label{fig:fxm}
\end{figure}

We calibrate the relation~(\ref{eq:xmass}) with and without the $z$-dependence 
using a subsample of ACT/MaDCoWS co-detections from \cite{2021A&A...653A.135O} with $z>0.7$ (see Section~\ref{sec:SZdata}). 
From both the source and background regions, we mask bright sources with the 0.5--2 keV flux above $10^{-13}$ ${ \rm erg\,s^{-1}\,cm^{-2}}$. Such sources are rare but introduce a large scatter to flux estimates. Additionally, in the background region, we mask all the detected sources down to 0.5--2 keV flux of $\sim2\times10^{-14} \,{ \rm erg\,s^{-1}\,cm^{-2}}$. We do not apply this flux cut to the source region since weak sources could not be unambiguously detected in the area dominated by a cluster emission (see a discussion in Appendix~\ref{sec:bias}). Therefore, the obtained source flux, corrected for the background contribution, needs to be further corrected for the excess in the resolved CXB fraction in the background region compared to the source aperture  (see column \texttt{`B1 (mean)}' for the sample of 36 objects in Table~\ref{fig:bias}).
We calculate flux uncertainties accounting for a Poisson noise   $\sigma_{\rm stat}$ (count statistics) and for the average variance in the background region in the 0.5--2 keV energy range (with the standard deviation $\sigma_{\rm var}=1.5\times10^{-14}$ erg s$^{-1}$ cm$^{-2}$). 
Figure~\ref{fig:overview} shows the resulting eROSITA fluxes in the 0.5--2~keV energy range,  masses $M_{\rm 500c}$, and redshifts for our subsample of clusters. 
The subsample of the \cite{2021A&A...653A.135O} catalog used here includes clusters with masses $\left (M_{\rm 500c}^{\rm UPP}\right )$ between $\sim 1.5\times 10^{14}M_{\odot}$ and  $\sim 6\times 10^{14}$ $M_{\odot}$ in the redshift range from $\sim 0.7$ to $\sim 1.4$. Corresponding fixed-aperture 0.5-2~keV X-ray fluxes vary from $\sim  10^{-14}$ erg s$^{-1}$ cm$^{-2}$ to $3\times 10^{-13}$ erg s$^{-1}$ cm$^{-2}$.

To fit the relation~\ref{eq:xmass} to the data\footnote{A more consistent approach would be writing a likelihood function that accounts for observational and intrinsic scatters and biases. However, for our approximate calibration, this is not required.}, we used the software package \texttt{ODRPACK} \citep{boggs1989user}, which implements the weighted Orthogonal Distance Regression and is designed to handle cases where both variables have measurement errors.
First, we consider $F_{\rm X}$--$M_{\rm 500c}$ relation without including a $z$ -dependence. As if there were no redshift estimates available except for knowing that clusters in the sample are at relatively high redshifts ($z \ge 0.5$).
As a result, we obtain $\eta = 0.81 \pm 0.04$, i.e. a $\sim$ 20\% difference in comparison with the $F_{\rm X}$--$M_{\rm 500c}$ relation from  \cite{2015MNRAS.450.1984C}.  As discussed above, particular methods of mass or flux estimation and the sample selection function may contribute to this difference.
Next, we fit the relation~(\ref{eq:xmass}) with the $z^{0.5}$-dependence to the data and obtain $\eta = 0.8 \pm 0.03$. As we see, including the $z^{0.5}$-dependence does not influence much the correction factor $\eta$. The redshift dependence was introduced to compensate for the trend in the X-ray flux-mass relation expected for the adopted scaling relation for clusters with masses larger than $10^{14}\Msun$. In practice, the total RMS scatter\footnote{calculated as a relative root mean square error = $\sqrt{\frac{1}{n}\sum{\left( \frac{M_{\rm X} - M^{\rm UPP}_{\rm 500c}}{M^{\rm UPP}_{\rm 500c}} \right)^2}}$} does not change much. It stays at the level of $\approx 34\%$. This means that the $z^{0.5}$ term can be omitted, i.e., it is sufficient to know that the cluster is beyond $z\sim 0.6$ and set $z=1$ to get the mass estimate with this level of accuracy.  

The resulting fit is shown in Fig.~\ref{fig:overview}. Dashed and dash-dotted lines show the relation~(\ref{eq:xmass}) with $\eta=1$ and 0.6, correspondingly. Figure~\ref{fig:fxm} shows the comparison between ACT masses and masses derived from the $F_{\rm X}$-$M_{\rm 500c}$ relation with $\eta = 0.8$.

If instead of $M_{\rm 500c}^{\rm UPP}$ masses we use  $M_{\rm 500c}^{\rm Cal} = M_{\rm 500c}^{\rm UPP}/0.71$ values, which are supposed to match weak-lensing masses on average  \citep[for details, see][]{2021ApJS..253....3H}, we obtain $\eta_{\rm WL}=\eta_{\rm UPP}/0.71\approx 1.13$.

While fitting the relation~(\ref{eq:xmass}), we do not exclude X-ray non-detections so as not to introduce an additional bias into the selection function. Excluding cluster candidates with no detectable X-ray emission lowers the best-fits value for $\eta$ but within the estimated uncertainties.

To sum up, the performed experiments with the ACT and eROSITA data suggest that for a pre-selected sample of high redshift galaxy clusters, one can estimate their masses from X-ray fluxes only with the RMS scatter in estimated masses ${\approx 34\%}$, even if a redshift is set to unity for all of them.
Several factors contribute to the scatter of the masses derived from $F_{\rm X}$ values relative to SZ-based masses in the  ACT/MaDCoWS subsample. Those include uncertainties in the SZ-based masses, photon-counting noise in the measured X-ray fluxes, and the intrinsic scatter in cluster properties. The value of the relative error in masses in the ACT/MaDCoWS sample is $\sim 0.17$ (on average). The intrinsic scatter in $\mathrm{ln}(L_{\rm X})$ for a given mass is $\sim$0.4 (based on the sample of well-observed clusters, e.g., \citealt{2009ApJ...692.1033V}) around the best-fitting relation $\mathrm{ln}(L_{\rm X})=1.61\, \mathrm{ln}(M)+const$. Assuming that (i) the same intrinsic scatter applies to clusters in the ACT/MADCoWS sample, and (ii) for estimates one can invert this relation to get the intrinsic scatter in $\mathrm{ln}(M)$ for a given $\mathrm{ln}(L_{\rm X})$, we get the expected intrinsic scatter in $\mathrm{ln}(M)$ for a given $\mathrm{ln}(L_{\rm X})$ of $\sim 0.4/1.61 \simeq 0.25$. Finally, the photon counting noise results in average $\sim 30\%$ uncertainty in $F_{\rm X}$  for the ACT/MADCoWS sample, which translates (using eq.~\ref{eq:CVS2015}) into a relative error in mass of the order of $0.3\times0.57\simeq 0.17$. The expected combined RMS deviation (assuming that these errors are independent) is at the level of $\sim 0.35$, which is close to the RMS scatter of 34\% obtained above.  We, therefore, believe that if the X-ray flux - mass relation (with $F_{\rm X}$ coming from all-sky surveys, i.e., available for any given position) can be used in combination with the optical/infrared data to impose an additional constraint (e.g., $F_{\rm X}$ greater than a certain value) to make the follow-up program more effective.  A construction of a fully Bayesian framework that allows combining all the available information to improve both samples’ purity and robustness of the mass estimates is straightforward to formulate but involves (unknown) pre-calibration factors and intrinsic scatter estimates, which can be achieved via comparison of high fidelity samples with low statistical noise and full-physics simulations encompassing large cosmological volumes \citep[e.g.,][]{2025arXiv250401061D}.   

Since the $F_{\rm X}$--$M_{\rm 500c}$ relation is applicable only to distant galaxy clusters, a pre-selection of such objects is needed. This can be facilitated, for example, by using photometric information from infrared and optical surveys, or by combining X-ray/SZ observations with upper limits on optical light from the brightest cluster galaxies, or by combining X-ray and SZ/radio observations \citep[e.g.,][]{2006MNRAS.373L..26V, 2016A&A...587A..23V, 2018AstL...44..297B,2021AstL...47..443B,2024A&A...690A.322K}.
The obtained $F_{\rm X}$-$M_{\rm 500c}$ relation can then be used, e.g., for the part of the MADCoWS sample for which no ACT mass estimates are available, and forced photometry of X-ray emission can be performed. An example of this application is given in \cite{2025A&A...695A.215D}, where the authors obtained 
masses for galaxy clusters with diffuse radio emission detectable in the LoTSS-DR2 data. Note, that \cite{2025A&A...695A.215D} used a slightly different value of $\eta$ coming from a similar calibration procedure but assuming that the ACT masses are known precisely.

Note that cosmological parameters used in this paper are slightly different from the ones used in \cite{2021A&A...653A.135O} and \cite{2021ApJS..253....3H}. Since we aim to connect a directly observable quantity ($F_{\rm X}$) to the cluster masses reported in these studies, one can simply interpret the resulting relation as a way of converting $F_{\rm X}$ to ACT masses.  It turns out that the uncertainty associated with such an estimate ($\approx 34\%$) is much larger than the one introduced by differences in the adopted cosmological parameters.

\subsection{$F_{\rm X}$-$M_{\rm 500c}$ applied to ACT DR5 clusters}

\begin{figure}
    \includegraphics[angle=0,trim=0cm 0cm 0.35cm 0cm,clip=true,width=0.9\columnwidth]{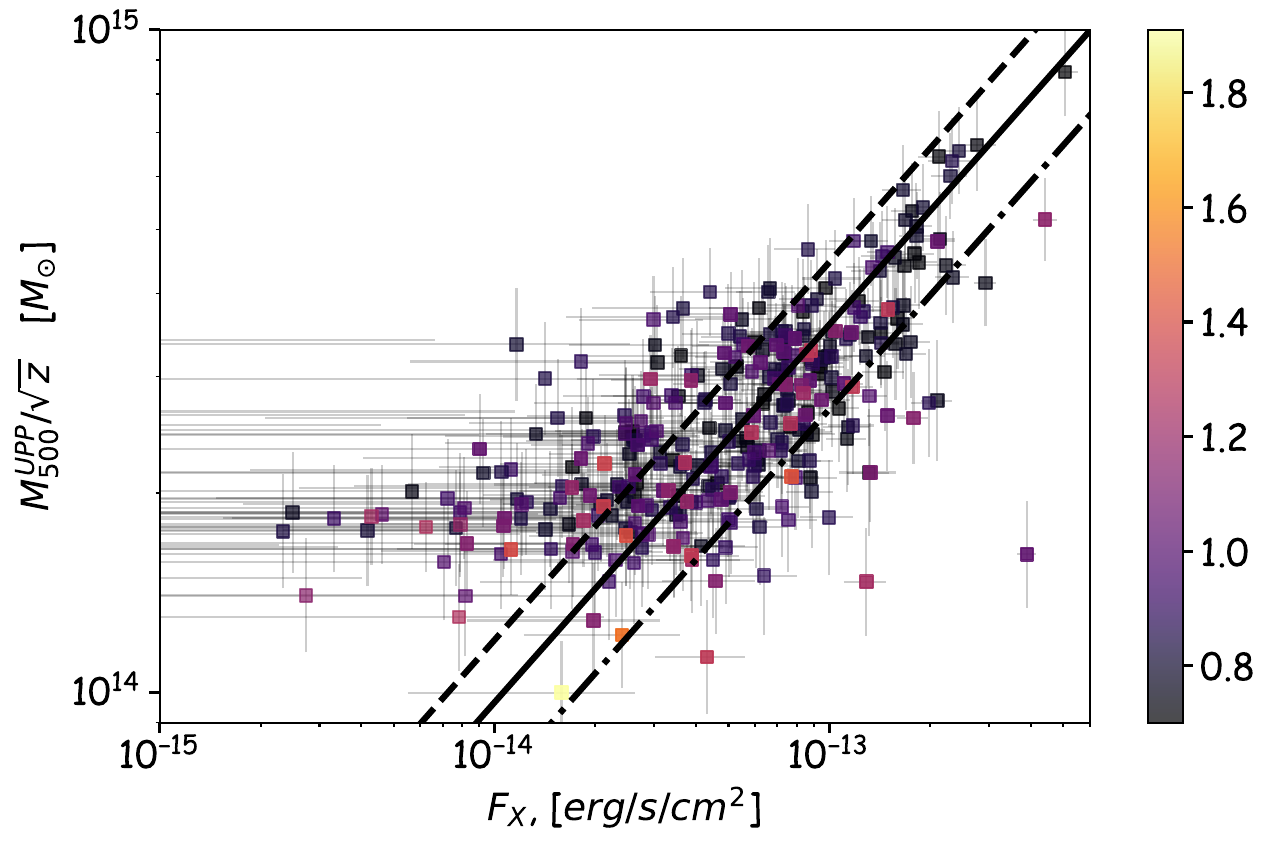}  
    \caption{ACT masses $M^{\rm UPP}_{\rm 500c}$ for the subsample of the ACT DR5 clusters with 0$^{\circ}$ < $l$ < 180$^{\circ}$ vs eROSITA 0.5-2.0 keV fluxes extracted from a circle of $R=2'$ arcmin centered at a cluster. The dashed, solid and dotted lines show the $F_{\rm X}$-$M_{\rm 500c}$ relation (equation~\ref{fig:fxm}) with $\eta = 1.0$, 0.8 and 0.6, correspondingly. Cluster redshifts are shown with color.}
    \label{fig:hilton}
\end{figure}

We illustrate the calibrated scaling relation between the X-ray flux and cluster mass on a larger ($\sim 400$ objects) sample of ACT DR5 clusters at $z>0.7$ (see Section~\ref{sec:SZdata}).
Source fluxes are estimated using the same procedure as in Section~\ref{sec:calib} but this time we do not mask sources in the source region and exclude only bright sources with flux above $ 10^{-13}$ erg s$^{-1}$ cm$^{-2}$ in the 0.5--2.0 keV band in the background region. The source flux is corrected for the expected bias according to Table~\ref{tab:bias}. The resulting flux estimates and the $F_{\rm X}$ - $M_{\rm 500c}$ relation with $\eta=$1 (dashed line), 0.8 (thick solid line), and 0.6 (dash-dotted line)  are shown in Fig.~\ref{fig:hilton}. While the bulk of clusters follow broadly the $F_{\rm X}$ - $M_{\rm 500c}$ scaling relation, there are several prominent outliers. The X-ray faint end of the distribution might contain a small fraction of false detections in SZ and be affected by uncertainties in X-ray measurements. Such objects have relatively low $M_{\rm 500c}^{\rm UPP}$ and X-ray fluxes consistent with zero.
Objects with very high X-ray flux (given the SZ mass estimate) appear to be clusters with an X-ray bright AGN in the center, clusters with very prominent cool cores, or with an unrelated bright X-ray source that happens to fall inside the source region.

We can also use the large sample of ACT DR5 clusters to estimate the scatter in mass at a given flux range. Since the X-ray data for the ACT DR5 sample were not cleaned for strong X-ray outliers (mostly due to the presence of AGN), we calculated 68\% quintiles of SZ-based masses around the $F_{\rm X}$-predicted values rather than the RMS. Furthermore, taking advantage of the larger sample,  we split the data in two X-ray flux bins: $3\times 10^{-14} \leq F_{\rm X} < 7.5 \times 10^{-14}\, {\rm erg\, s^{-1}\, cm^{-2}}$ and $F_{\rm X} \ge 7.5 \times 10^{-14}\,{\rm erg\,s^{-1}\,cm^{-2}}$. Clusters with X-ray fluxes below $3\times 10^{-14} \, {\rm erg\, s^{-1}\, cm^{-2}}$ were excluded to avoid too strong contribution of the photon counting noise. Two flux ranges were chosen to contain roughly the same number of objects (124/113 galaxy clusters in lower/higher flux bins). Approximately 68\% of ACT masses ($M_{\rm 500c}^{\rm UPP}$) lie within $\pm 24$\% of the $F_{\rm X}$-based estimates for lower X-ray fluxes, and $\pm 23$\% for higher X-ray fluxes.

Both X-ray \citep{2009ApJ...692.1033V} and SZ \citep{2021A&A...653A.135O, 2021ApJS..253....3H} samples cover approximately the same mass range $M \gtrsim 1.5 \times 10^{14} M_\odot$, but the SZ-based sample extends to higher redshifts. Therefore, our analysis confirms that $F_{\rm X}$ remains a crude indicator of cluster masses for higher mean redshifts than the original X-ray sample of  \cite{2009ApJ...692.1033V}. Note that there is no guarantee that an extrapolation to lower masses (groups of galaxies) holds, too. In particular, at lower temperatures, the X-ray emissivity depends strongly on the abundance of metals and varies non-monotonically with temperature (see, e.g.,  Fig.B2 in \citealt{2023MNRAS.525..898L}).

\section*{Conclusions}
Our results can be summarized as follows:
\begin{itemize}
\item Using a combination of eROSITA X-ray data and SZ-based measurements of galaxy cluster masses from ACT samples, we confirmed that observed X-ray flux (rather than luminosity) in the 0.5-2~keV band can be used as a direct proxy for cluster masses at $z\gtrsim 0.6-0.7$ in the form
\begin{equation}
   M_{\rm 500c} =    1.2\times 10^{14} M_\odot\; \eta\;\left ( \frac{F_{\rm X}}{10^{-14}\,{\rm erg\,s^{-1}cm^{-2}}}\right )^{0.57}\;z^{0.5} 
\end{equation}
suggested in \citealt{2015MNRAS.450.1984C} to which we have added a correction factor $z^{0.5}$ to extend the applicability range. 

This relation is based on the extrapolation of the X-ray-based scaling relations from \citealt{2009ApJ...692.1033V} and the standard $\Lambda$CDM model.
\item Using the sample of \citealt{2021A&A...653A.135O}, which consists of clusters detected both in the  MaDCoWS \citep{2019ApJS..240...33G} and ACT data, we obtained  $\eta_{\rm UPP}\approx 0.80$ if $M_{\rm 500c}^{\rm UPP}$ is used. 
The resulting RMS deviation of the masses predicted by the best-fitting $F_{\rm X}$--$M_{\rm 500c}$ relation from the ACT masses is $\approx 34\%$. The masses recalibrated to match weak lensing analysis $\left (M_{\rm 500c}^{\rm Cal}\right )$ from \citealt{2021ApJS..253....3H} are related as $M_{\rm 500c}^{\rm UPP}=0.71\,M_{\rm 500c}^{\rm Cal}$. Therefore, if $M_{\rm 500c}^{\rm Cal}$ is used, $\eta_{\rm WL}=\eta_{\rm UPP}/0.71\approx 1.13$. These results suggest that the X-ray-based scaling relations predict masses halfway between SZ and weak lensing mass calibration.
\item Overall, the relation of the X-ray flux and mass can be used even without knowing the exact redshift, i.e., setting $z=1$, to get a rough estimate of cluster masses (as long as it is known that $z\gtrsim 0.6-0.7$). It is particularly useful for a pre-selection of massive cluster candidates (for follow-up verification).   
\end{itemize}

\begin{acknowledgements}
We are very grateful to the referee for a thorough reading of the manuscript and  for the valuable comments.
This work is based on observations with eROSITA telescope onboard SRG observatory. The SRG observatory was built by Roskosmos in the interests of the Russian Academy of Sciences represented by its Space Research Institute (IKI) in the framework of the Russian Federal Space Program, with the participation of the Deutsches Zentrum für Luft- und Raumfahrt (DLR). The SRG/eROSITA X-ray telescope was built by a consortium of German Institutes led by MPE, and supported by DLR.  The SRG spacecraft was designed, built, launched and is operated by the Lavochkin Association and its subcontractors. The science data are downlinked via the Deep Space Network Antennae in Bear Lakes, Ussurijsk, and Baykonur, funded by Roskosmos. The eROSITA data used in this work were processed using the eSASS software system developed by the German eROSITA consortium and proprietary data reduction and analysis software developed by the Russian eROSITA Consortium.

IK acknowledges support by the COMPLEX project from the European Research Council (ERC) under the European Union’s Horizon 2020 research and innovation program grant agreement ERC-2019-AdG 882679.

NL acknowledges financial support from the Russian Science Foundation (grant no. 25-22-00470).

\end{acknowledgements}

\bibliographystyle{aa}

\bibliography{ref}

\begin{thebibliography}{45}
\expandafter\ifx\csname natexlab\endcsname\relax\def\natexlab#1{#1}\fi

\bibitem[{{Allen} {et~al.}(2011){Allen}, {Evrard}, \&
  {Mantz}}]{2011ARA&A..49..409A}
{Allen}, S.~W., {Evrard}, A.~E., \& {Mantz}, A.~B. 2011, \araa, 49, 409

\bibitem[{{Arnaud} {et~al.}(2010){Arnaud}, {Pratt}, {Piffaretti},
  {B{\"o}hringer}, {Croston}, \& {Pointecouteau}}]{2010A&A...517A..92A}
{Arnaud}, M., {Pratt}, G.~W., {Piffaretti}, R., {et~al.} 2010, \aap, 517, A92

\bibitem[{Boggs {et~al.}(1989)Boggs, Byrd, Donaldson, \&
  Schnabel}]{boggs1989user}
Boggs, P., Byrd, R., Donaldson, J., \& Schnabel, R. 1989

\bibitem[{{Brunner} {et~al.}(2018){Brunner}, {Boller}, {Coutinho}, {Dauser},
  {Dennerl}, {Dwelly}, {Freyberg}, {F{\"u}rmetz}, {Georgakakis}, {Grossberger},
  {Kreykenbohm}, {Lamer}, {Meidinger}, {M{\"u}ller}, {Predehl}, {Robrade},
  {Sanders}, \& {Wilms}}]{2018SPIE10699E..5GB}
{Brunner}, H., {Boller}, T., {Coutinho}, D., {et~al.} 2018, in Society of
  Photo-Optical Instrumentation Engineers (SPIE) Conference Series, Vol. 10699,
  Space Telescopes and Instrumentation 2018: Ultraviolet to Gamma Ray, ed.
  J.-W.~A. {den Herder}, S.~{Nikzad}, \& K.~{Nakazawa}, 106995G

\bibitem[{{Burenin} {et~al.}(2021){Burenin}, {Bikmaev}, {Gilfanov},
  {Grokhovskaya}, {Dodonov}, {Eselevich}, {Zaznobin}, {Irtuganov}, {Lyskova},
  {Medvedev}, {Meshcheryakov}, {Moiseev}, {Sazonov}, {Starobinsky}, {Sunyaev},
  {Uklein}, {Khabibullin}, {Khamitov}, \& {Churazov}}]{2021AstL...47..443B}
{Burenin}, R.~A., {Bikmaev}, I.~F., {Gilfanov}, M.~R., {et~al.} 2021, Astronomy
  Letters, 47, 443

\bibitem[{{Burenin} {et~al.}(2018){Burenin}, {Bikmaev}, {Khamitov}, {Zaznobin},
  {Khorunzhev}, {Eselevich}, {Afanasiev}, {Dodonov}, {Rubi{\~n}o-Mart{\'\i}n},
  {Aghanim}, \& {Sunyaev}}]{2018AstL...44..297B}
{Burenin}, R.~A., {Bikmaev}, I.~F., {Khamitov}, I.~M., {et~al.} 2018, Astronomy
  Letters, 44, 297

\bibitem[{{Churazov} {et~al.}(2021){Churazov}, {Khabibullin}, {Lyskova},
  {Sunyaev}, \& {Bykov}}]{2021A&A...651A..41C}
{Churazov}, E., {Khabibullin}, I., {Lyskova}, N., {Sunyaev}, R., \& {Bykov},
  A.~M. 2021, \aap, 651, A41

\bibitem[{{Churazov} {et~al.}(2015){Churazov}, {Vikhlinin}, \&
  {Sunyaev}}]{2015MNRAS.450.1984C}
{Churazov}, E., {Vikhlinin}, A., \& {Sunyaev}, R. 2015, \mnras, 450, 1984

\bibitem[{{Coulton} {et~al.}(2024){Coulton}, {Madhavacheril}, {Duivenvoorden},
  {Hill}, {Abril-Cabezas}, {Ade}, {Aiola}, {Alford}, {Amiri}, {Amodeo}, {An},
  {Atkins}, {Austermann}, {Battaglia}, {Battistelli}, {Beall}, {Bean},
  {Beringue}, {Bhandarkar}, {Biermann}, {Bolliet}, {Bond}, {Cai}, {Calabrese},
  {Calafut}, {Capalbo}, {Carrero}, {Chesmore}, {Cho}, {Choi}, {Clark},
  {Rosado}, {Cothard}, {Coughlin}, {Crowley}, {Devlin}, {Dicker}, {Doze},
  {Duell}, {Duff}, {Dunkley}, {D{\"u}nner}, {Fanfani}, {Fankhanel}, {Farren},
  {Ferraro}, {Freundt}, {Fuzia}, {Gallardo}, {Garrido}, {Givans}, {Gluscevic},
  {Golec}, {Guan}, {Halpern}, {Han}, {Hasselfield}, {Healy}, {Henderson},
  {Hensley}, {Herv{\'\i}as-Caimapo}, {Hilton}, {Hilton}, {Hincks},
  {Hlo{\v{z}}ek}, {Ho}, {Huber}, {Hubmayr}, {Huffenberger}, {Hughes}, {Irwin},
  {Isopi}, {Jense}, {Keller}, {Kim}, {Knowles}, {Koopman}, {Kosowsky},
  {Kramer}, {Kusiak}, {La Posta}, {Lakey}, {Lee}, {Li}, {Li}, {Limon},
  {Lokken}, {Louis}, {Lungu}, {MacCrann}, {MacInnis}, {Maldonado}, {Maldonado},
  {Mallaby-Kay}, {Marques}, {van Marrewijk}, {McCarthy}, {McMahon}, {Mehta},
  {Menanteau}, {Moodley}, {Morris}, {Mroczkowski}, {Naess}, {Namikawa}, {Nati},
  {Newburgh}, {Nicola}, {Niemack}, {Nolta}, {Orlowski-Scherer}, {Page},
  {Pandey}, {Partridge}, {Prince}, {Puddu}, {Qu}, {Radiconi}, {Robertson},
  {Rojas}, {Sakuma}, {Salatino}, {Schaan}, {Schmitt}, {Sehgal}, {Shaikh},
  {Sherwin}, {Sierra}, {Sievers}, {Sif{\'o}n}, {Simon}, {Sonka}, {Spergel},
  {Staggs}, {Storer}, {Switzer}, {Tampier}, {Thornton}, {Trac}, {Treu},
  {Tucker}, {Ullom}, {Vale}, {Van Engelen}, {Van Lanen}, {Vargas},
  {Vavagiakis}, {Wagoner}, {Wang}, {Wenzl}, {Wollack}, {Xu}, {Zago}, \&
  {Zheng}}]{2024PhRvD.109f3530C}
{Coulton}, W., {Madhavacheril}, M.~S., {Duivenvoorden}, A.~J., {et~al.} 2024,
  \prd, 109, 063530

\bibitem[{{DESI Collaboration} {et~al.}(2016){DESI Collaboration}, {Aghamousa},
  {Aguilar}, {Ahlen}, {Alam}, {Allen}, {Allende Prieto}, {Annis}, {Bailey},
  {Balland}, {Ballester}, {Baltay}, {Beaufore}, {Bebek}, {Beers}, {Bell},
  {Bernal}, {Besuner}, {Beutler}, {Blake}, {Bleuler}, {Blomqvist}, {Blum},
  {Bolton}, {Briceno}, {Brooks}, {Brownstein}, {Buckley-Geer}, {Burden},
  {Burtin}, {Busca}, {Cahn}, {Cai}, {Cardiel-Sas}, {Carlberg}, {Carton},
  {Casas}, {Castander}, {Cervantes-Cota}, {Claybaugh}, {Close}, {Coker},
  {Cole}, {Comparat}, {Cooper}, {Cousinou}, {Crocce}, {Cuby}, {Cunningham},
  {Davis}, {Dawson}, {de la Macorra}, {De Vicente}, {Delubac}, {Derwent},
  {Dey}, {Dhungana}, {Ding}, {Doel}, {Duan}, {Ealet}, {Edelstein},
  {Eftekharzadeh}, {Eisenstein}, {Elliott}, {Escoffier}, {Evatt}, {Fagrelius},
  {Fan}, {Fanning}, {Farahi}, {Farihi}, {Favole}, {Feng}, {Fernandez},
  {Findlay}, {Finkbeiner}, {Fitzpatrick}, {Flaugher}, {Flender}, {Font-Ribera},
  {Forero-Romero}, {Fosalba}, {Frenk}, {Fumagalli}, {Gaensicke}, {Gallo},
  {Garcia-Bellido}, {Gaztanaga}, {Pietro Gentile Fusillo}, {Gerard},
  {Gershkovich}, {Giannantonio}, {Gillet}, {Gonzalez-de-Rivera},
  {Gonzalez-Perez}, {Gott}, {Graur}, {Gutierrez}, {Guy}, {Habib}, {Heetderks},
  {Heetderks}, {Heitmann}, {Hellwing}, {Herrera}, {Ho}, {Holland}, {Honscheid},
  {Huff}, {Hutchinson}, {Huterer}, {Hwang}, {Illa Laguna}, {Ishikawa},
  {Jacobs}, {Jeffrey}, {Jelinsky}, {Jennings}, {Jiang}, {Jimenez}, {Johnson},
  {Joyce}, {Jullo}, {Juneau}, {Kama}, {Karcher}, {Karkar}, {Kehoe}, {Kennamer},
  {Kent}, {Kilbinger}, {Kim}, {Kirkby}, {Kisner}, {Kitanidis}, {Kneib},
  {Koposov}, {Kovacs}, {Koyama}, {Kremin}, {Kron}, {Kronig}, {Kueter-Young},
  {Lacey}, {Lafever}, {Lahav}, {Lambert}, {Lampton}, {Landriau}, {Lang},
  {Lauer}, {Le Goff}, {Le Guillou}, {Le Van Suu}, {Lee}, {Lee}, {Leitner},
  {Lesser}, {Levi}, {L'Huillier}, {Li}, {Liang}, {Lin}, {Linder}, {Loebman},
  {Luki{\'c}}, {Ma}, {MacCrann}, {Magneville}, {Makarem}, {Manera}, {Manser},
  {Marshall}, {Martini}, {Massey}, {Matheson}, {McCauley}, {McDonald},
  {McGreer}, {Meisner}, {Metcalfe}, {Miller}, {Miquel}, {Moustakas}, {Myers},
  {Naik}, {Newman}, {Nichol}, {Nicola}, {Nicolati da Costa}, {Nie}, {Niz},
  {Norberg}, {Nord}, {Norman}, {Nugent}, {O'Brien}, {Oh}, \&
  {Olsen}}]{2016arXiv161100036D}
{DESI Collaboration}, {Aghamousa}, A., {Aguilar}, J., {et~al.} 2016, arXiv
  e-prints, arXiv:1611.00036

\bibitem[{{Di Gennaro} {et~al.}(2025){Di Gennaro}, {Br{\"u}ggen}, {Moravec},
  {Di Mascolo}, {van Weeren}, {Brunetti}, {Cassano}, {Botteon}, {Churazov},
  {Khabibullin}, {Lyskova}, {de Gasperin}, {Hardcastle}, {R{\"o}ttgering},
  {Shimwell}, {Sunyaev}, \& {Stanford}}]{2025A&A...695A.215D}
{Di Gennaro}, G., {Br{\"u}ggen}, M., {Moravec}, E., {et~al.} 2025, \aap, 695,
  A215

\bibitem[{{Dolag} {et~al.}(2025){Dolag}, {Remus}, {Valenzuela}, {Kimmig},
  {Seidel}, {Fortune}, {Stoiber}, {Ivleva}, {Hoffmann}, {Biffi}, {Marini},
  {Popesso}, \& {Vladutescu-Zopp}}]{2025arXiv250401061D}
{Dolag}, K., {Remus}, R.-S., {Valenzuela}, L.~M., {et~al.} 2025, arXiv
  e-prints, arXiv:2504.01061

\bibitem[{{Gilfanov} {et~al.}(2004){Gilfanov}, {Grimm}, \&
  {Sunyaev}}]{2004MNRAS.351.1365G}
{Gilfanov}, M., {Grimm}, H.~J., \& {Sunyaev}, R. 2004, \mnras, 351, 1365

\bibitem[{{Giodini} {et~al.}(2013){Giodini}, {Lovisari}, {Pointecouteau},
  {Ettori}, {Reiprich}, \& {Hoekstra}}]{2013SSRv..177..247G}
{Giodini}, S., {Lovisari}, L., {Pointecouteau}, E., {et~al.} 2013, \ssr, 177,
  247

\bibitem[{{Gonzalez} {et~al.}(2019){Gonzalez}, {Gettings}, {Brodwin},
  {Eisenhardt}, {Stanford}, {Wylezalek}, {Decker}, {Marrone}, {Moravec},
  {O'Donnell}, {Stalder}, {Stern}, {Abdulla}, {Brown}, {Carlstrom}, {Chambers},
  {Hayden}, {Lin}, {Magnier}, {Masci}, {Mantz}, {McDonald}, {Mo}, {Perlmutter},
  {Wright}, \& {Zeimann}}]{2019ApJS..240...33G}
{Gonzalez}, A.~H., {Gettings}, D.~P., {Brodwin}, M., {et~al.} 2019, \apjs, 240,
  33

\bibitem[{{Hilton} {et~al.}(2021){Hilton}, {Sif{\'o}n}, {Naess},
  {Madhavacheril}, {Oguri}, {Rozo}, {Rykoff}, {Abbott}, {Adhikari}, {Aguena},
  {Aiola}, {Allam}, {Amodeo}, {Amon}, {Annis}, {Ansarinejad}, {Aros-Bunster},
  {Austermann}, {Avila}, {Bacon}, {Battaglia}, {Beall}, {Becker}, {Bernstein},
  {Bertin}, {Bhandarkar}, {Bhargava}, {Bond}, {Brooks}, {Burke}, {Calabrese},
  {Carrasco Kind}, {Carretero}, {Choi}, {Choi}, {Conselice}, {da Costa},
  {Costanzi}, {Crichton}, {Crowley}, {D{\"u}nner}, {Denison}, {Devlin},
  {Dicker}, {Diehl}, {Dietrich}, {Doel}, {Duff}, {Duivenvoorden}, {Dunkley},
  {Everett}, {Ferraro}, {Ferrero}, {Fert{\'e}}, {Flaugher}, {Frieman},
  {Gallardo}, {Garc{\'\i}a-Bellido}, {Gaztanaga}, {Gerdes}, {Giles}, {Golec},
  {Gralla}, {Grandis}, {Gruen}, {Gruendl}, {Gschwend}, {Gutierrez}, {Han},
  {Hartley}, {Hasselfield}, {Hill}, {Hilton}, {Hincks}, {Hinton}, {Ho},
  {Honscheid}, {Hoyle}, {Hubmayr}, {Huffenberger}, {Hughes}, {Jaelani}, {Jain},
  {James}, {Jeltema}, {Kent}, {Knowles}, {Koopman}, {Kuehn}, {Lahav}, {Lima},
  {Lin}, {Lokken}, {Loubser}, {MacCrann}, {Maia}, {Marriage}, {Martin},
  {McMahon}, {Melchior}, {Menanteau}, {Miquel}, {Miyatake}, {Moodley},
  {Morgan}, {Mroczkowski}, {Nati}, {Newburgh}, {Niemack}, {Nishizawa},
  {Ogando}, {Orlowski-Scherer}, {Page}, {Palmese}, {Partridge},
  {Paz-Chinch{\'o}n}, {Phakathi}, {Plazas}, {Robertson}, {Romer}, {Carnero
  Rosell}, {Salatino}, {Sanchez}, {Schaan}, {Schillaci}, {Sehgal}, {Serrano},
  {Shin}, {Simon}, {Smith}, {Soares-Santos}, {Spergel}, {Staggs}, {Storer},
  {Suchyta}, {Swanson}, {Tarle}, {Thomas}, {To}, {Trac}, {Ullom}, {Vale}, {Van
  Lanen}, {Vavagiakis}, {De Vicente}, {Wilkinson}, {Wollack}, {Xu}, \&
  {Zhang}}]{2021ApJS..253....3H}
{Hilton}, M., {Sif{\'o}n}, C., {Naess}, S., {et~al.} 2021, \apjs, 253, 3

\bibitem[{{Jenkins} {et~al.}(2001){Jenkins}, {Frenk}, {White}, {Colberg},
  {Cole}, {Evrard}, {Couchman}, \& {Yoshida}}]{2001MNRAS.321..372J}
{Jenkins}, A., {Frenk}, C.~S., {White}, S.~D.~M., {et~al.} 2001, \mnras, 321,
  372

\bibitem[{{Khabibullin} {et~al.}(2023){Khabibullin}, {Churazov}, {Bykov},
  {Chugai}, \& {Sunyaev}}]{2023MNRAS.521.5536K}
{Khabibullin}, I.~I., {Churazov}, E.~M., {Bykov}, A.~M., {Chugai}, N.~N., \&
  {Sunyaev}, R.~A. 2023, \mnras, 521, 5536

\bibitem[{{Klein} {et~al.}(2024){Klein}, {Mohr}, \&
  {Davies}}]{2024A&A...690A.322K}
{Klein}, M., {Mohr}, J.~J., \& {Davies}, C.~T. 2024, \aap, 690, A322

\bibitem[{{Kravtsov} \& {Borgani}(2012)}]{2012ARA&A..50..353K}
{Kravtsov}, A.~V. \& {Borgani}, S. 2012, \araa, 50, 353

\bibitem[{{Kravtsov} {et~al.}(2006){Kravtsov}, {Vikhlinin}, \&
  {Nagai}}]{2006ApJ...650..128K}
{Kravtsov}, A.~V., {Vikhlinin}, A., \& {Nagai}, D. 2006, \apj, 650, 128

\bibitem[{{Kruglov} {et~al.}(2025){Kruglov}, {Khabibullin}, {Lyskova}, {Dolag},
  \& {Biffi}}]{2024arXiv240812026K}
{Kruglov}, A., {Khabibullin}, I., {Lyskova}, N., {Dolag}, K., \& {Biffi}, V.
  2025, \jcap, 2025, 007

\bibitem[{{Luo} {et~al.}(2017){Luo}, {Brandt}, {Xue}, {Lehmer}, {Alexander},
  {Bauer}, {Vito}, {Yang}, {Basu-Zych}, {Comastri}, {Gilli}, {Gu},
  {Hornschemeier}, {Koekemoer}, {Liu}, {Mainieri}, {Paolillo}, {Ranalli},
  {Rosati}, {Schneider}, {Shemmer}, {Smail}, {Sun}, {Tozzi}, {Vignali}, \&
  {Wang}}]{2017ApJS..228....2L}
{Luo}, B., {Brandt}, W.~N., {Xue}, Y.~Q., {et~al.} 2017, \apjs, 228, 2

\bibitem[{{Lyskova} {et~al.}(2023){Lyskova}, {Churazov}, {Khabibullin},
  {Burenin}, {Starobinsky}, \& {Sunyaev}}]{2023MNRAS.525..898L}
{Lyskova}, N., {Churazov}, E., {Khabibullin}, I.~I., {et~al.} 2023, \mnras,
  525, 898

\bibitem[{{Mantz} {et~al.}(2018){Mantz}, {Allen}, {Morris}, \& {von der
  Linden}}]{2018MNRAS.473.3072M}
{Mantz}, A.~B., {Allen}, S.~W., {Morris}, R.~G., \& {von der Linden}, A. 2018,
  \mnras, 473, 3072

\bibitem[{{Mantz} {et~al.}(2016){Mantz}, {Allen}, {Morris}, {von der Linden},
  {Applegate}, {Kelly}, {Burke}, {Donovan}, \& {Ebeling}}]{2016MNRAS.463.3582M}
{Mantz}, A.~B., {Allen}, S.~W., {Morris}, R.~G., {et~al.} 2016, \mnras, 463,
  3582

\bibitem[{{Massey} {et~al.}(2013){Massey}, {Hoekstra}, {Kitching}, {Rhodes},
  {Cropper}, {Amiaux}, {Harvey}, {Mellier}, {Meneghetti}, {Miller},
  {Paulin-Henriksson}, {Pires}, {Scaramella}, \&
  {Schrabback}}]{2013MNRAS.429..661M}
{Massey}, R., {Hoekstra}, H., {Kitching}, T., {et~al.} 2013, \mnras, 429, 661

\bibitem[{{Nagai}(2006)}]{2006ApJ...650..538N}
{Nagai}, D. 2006, \apj, 650, 538

\bibitem[{{Oguri} \& {Miyazaki}(2025)}]{2025PJAB..101..129O}
{Oguri}, M. \& {Miyazaki}, S. 2025, Proceedings of the Japan Academy, Series B,
  101, 129

\bibitem[{{Orlowski-Scherer} {et~al.}(2021){Orlowski-Scherer}, {Di Mascolo},
  {Bhandarkar}, {Manduca}, {Mroczkowski}, {Amodeo}, {Battaglia}, {Brodwin},
  {Choi}, {Devlin}, {Dicker}, {Dunkley}, {Gonzalez}, {Han}, {Hilton},
  {Huffenberger}, {Hughes}, {MacInnis}, {Knowles}, {Koopman}, {Lowe},
  {Moodley}, {Nati}, {Niemack}, {Page}, {Partridge}, {Romero}, {Salatino},
  {Schillaci}, {Sehgal}, {Sif{\'o}n}, {Staggs}, {Stanford}, {Thornton},
  {Vavagiakis}, {Wollack}, {Xu}, \& {Zhu}}]{2021A&A...653A.135O}
{Orlowski-Scherer}, J., {Di Mascolo}, L., {Bhandarkar}, T., {et~al.} 2021,
  \aap, 653, A135

\bibitem[{{Planck Collaboration} {et~al.}(2013){Planck Collaboration}, {Ade},
  {Aghanim}, {Arnaud}, {Ashdown}, {Atrio-Barandela}, {Aumont}, {Baccigalupi},
  {Balbi}, {Banday}, {Barreiro}, {Bartlett}, {Battaner}, {Battye}, {Benabed},
  {Bernard}, {Bersanelli}, {Bhatia}, {Bikmaev}, {B{\"o}hringer}, {Bonaldi},
  {Bond}, {Borgani}, {Borrill}, {Bouchet}, {Bourdin}, {Brown}, {Bucher},
  {Burenin}, {Burigana}, {Butler}, {Cabella}, {Cardoso}, {Carvalho},
  {Chamballu}, {Chiang}, {Chon}, {Clements}, {Colafrancesco}, {Coulais},
  {Cuttaia}, {Da Silva}, {Dahle}, {Davis}, {de Bernardis}, {de Gasperis},
  {Delabrouille}, {D{\'e}mocl{\`e}s}, {D{\'e}sert}, {Diego}, {Dolag}, {Dole},
  {Donzelli}, {Dor{\'e}}, {Douspis}, {Dupac}, {Efstathiou}, {En{\ss}lin},
  {Eriksen}, {Finelli}, {Flores-Cacho}, {Forni}, {Frailis}, {Franceschi},
  {Frommert}, {Galeotta}, {Ganga}, {G{\'e}nova-Santos}, {Giard},
  {Giraud-H{\'e}raud}, {Gonz{\'a}lez-Nuevo}, {G{\'o}rski}, {Gregorio},
  {Gruppuso}, {Hansen}, {Harrison}, {Hern{\'a}ndez-Monteagudo}, {Herranz},
  {Hildebrandt}, {Hivon}, {Hobson}, {Holmes}, {Huffenberger}, {Hurier},
  {Jagemann}, {Juvela}, {Keih{\"a}nen}, {Khamitov}, {Kneissl}, {Knoche},
  {Kunz}, {Kurki-Suonio}, {Lagache}, {Lamarre}, {Lasenby}, {Lawrence}, {Le
  Jeune}, {Leach}, {Leonardi}, {Liddle}, {Lilje}, {Linden-V{\o}rnle},
  {L{\'o}pez-Caniego}, {Luzzi}, {Mac{\'\i}as-P{\'e}rez}, {Maino}, {Mandolesi},
  {Maris}, {Marleau}, {Marshall}, {Mart{\'\i}nez-Gonz{\'a}lez}, {Masi},
  {Matarrese}, {Matthai}, {Mazzotta}, {Meinhold}, {Melchiorri}, {Melin},
  {Mendes}, {Mitra}, {Miville-Desch{\^e}nes}, {Montier}, {Morgante}, {Munshi},
  {Natoli}, {N{\o}rgaard-Nielsen}, {Noviello}, {Osborne}, {Pajot}, {Paoletti},
  {Partridge}, {Pearson}, {Perdereau}, {Perrotta}, {Piacentini}, {Piat},
  {Pierpaoli}, {Piffaretti}, {Platania}, {Pointecouteau}, {Polenta},
  {Ponthieu}, {Popa}, {Poutanen}, {Pratt}, {Prunet}, {Puget}, {Rachen},
  {Rebolo}, {Reinecke}, {Remazeilles}, {Renault}, {Ricciardi}, {Ristorcelli},
  {Rocha}, {Rosset}, {Rossetti}, {Rubi{\~n}o-Mart{\'\i}n}, {Rusholme},
  {Sandri}, {Savini}, {Scott}, {Starck}, {Stivoli}, {Stolyarov}, {Sudiwala},
  {Sunyaev}, {Sutton}, {Suur-Uski}, {Sygnet}, {Tauber}, {Terenzi},
  {Toffolatti}, {Tomasi}, {Tristram}, {Valenziano}, {Van Tent}, {Vielva},
  {Villa}, {Vittorio}, {Wandelt}, {Weller}, {White}, {Yvon}, {Zacchei}, \&
  {Zonca}}]{2013A&A...550A.129P}
{Planck Collaboration}, {Ade}, P.~A.~R., {Aghanim}, N., {et~al.} 2013, \aap,
  550, A129

\bibitem[{{Pratt} {et~al.}(2019){Pratt}, {Arnaud}, {Biviano}, {Eckert},
  {Ettori}, {Nagai}, {Okabe}, \& {Reiprich}}]{2019SSRv..215...25P}
{Pratt}, G.~W., {Arnaud}, M., {Biviano}, A., {et~al.} 2019, \ssr, 215, 25

\bibitem[{{Pratt} {et~al.}(2022){Pratt}, {Arnaud}, {Maughan}, \&
  {Melin}}]{2022A&A...665A..24P}
{Pratt}, G.~W., {Arnaud}, M., {Maughan}, B.~J., \& {Melin}, J.~B. 2022, \aap,
  665, A24

\bibitem[{{Pratt} {et~al.}(2009){Pratt}, {Croston}, {Arnaud}, \&
  {B{\"o}hringer}}]{2009A&A...498..361P}
{Pratt}, G.~W., {Croston}, J.~H., {Arnaud}, M., \& {B{\"o}hringer}, H. 2009,
  \aap, 498, 361

\bibitem[{{Predehl} {et~al.}(2021){Predehl}, {Andritschke}, {Arefiev},
  {Babyshkin}, {Batanov}, {Becker}, {B{\"o}hringer}, {Bogomolov}, {Boller},
  {Borm}, {Bornemann}, {Br{\"a}uninger}, {Br{\"u}ggen}, {Brunner}, {Brusa},
  {Bulbul}, {Buntov}, {Burwitz}, {Burkert}, {Clerc}, {Churazov}, {Coutinho},
  {Dauser}, {Dennerl}, {Doroshenko}, {Eder}, {Emberger}, {Eraerds},
  {Finoguenov}, {Freyberg}, {Friedrich}, {Friedrich}, {F{\"u}rmetz},
  {Georgakakis}, {Gilfanov}, {Granato}, {Grossberger}, {Gueguen}, {Gureev},
  {Haberl}, {H{\"a}lker}, {Hartner}, {Hasinger}, {Huber}, {Ji}, {Kienlin},
  {Kink}, {Korotkov}, {Kreykenbohm}, {Lamer}, {Lomakin}, {Lapshov}, {Liu},
  {Maitra}, {Meidinger}, {Menz}, {Merloni}, {Mernik}, {Mican}, {Mohr},
  {M{\"u}ller}, {Nandra}, {Nazarov}, {Pacaud}, {Pavlinsky}, {Perinati},
  {Pfeffermann}, {Pietschner}, {Ramos-Ceja}, {Rau}, {Reiffers}, {Reiprich},
  {Robrade}, {Salvato}, {Sanders}, {Santangelo}, {Sasaki}, {Scheuerle},
  {Schmid}, {Schmitt}, {Schwope}, {Shirshakov}, {Steinmetz}, {Stewart},
  {Str{\"u}der}, {Sunyaev}, {Tenzer}, {Tiedemann}, {Tr{\"u}mper}, {Voron},
  {Weber}, {Wilms}, \& {Yaroshenko}}]{2021A&A...647A...1P}
{Predehl}, P., {Andritschke}, R., {Arefiev}, V., {et~al.} 2021, \aap, 647, A1

\bibitem[{{Sheth} {et~al.}(2001){Sheth}, {Mo}, \&
  {Tormen}}]{2001MNRAS.323....1S}
{Sheth}, R.~K., {Mo}, H.~J., \& {Tormen}, G. 2001, \mnras, 323, 1

\bibitem[{{Sunyaev} {et~al.}(2021){Sunyaev}, {Arefiev}, {Babyshkin},
  {Bogomolov}, {Borisov}, {Buntov}, {Brunner}, {Burenin}, {Churazov},
  {Coutinho}, {Eder}, {Eismont}, {Freyberg}, {Gilfanov}, {Gureyev}, {Hasinger},
  {Khabibullin}, {Kolmykov}, {Komovkin}, {Krivonos}, {Lapshov}, {Levin},
  {Lomakin}, {Lutovinov}, {Medvedev}, {Merloni}, {Mernik}, {Mikhailov},
  {Molodtsov}, {Mzhelsky}, {M{\"u}ller}, {Nandra}, {Nazarov}, {Pavlinsky},
  {Poghodin}, {Predehl}, {Robrade}, {Sazonov}, {Scheuerle}, {Shirshakov},
  {Tkachenko}, \& {Voron}}]{2021A&A...656A.132S}
{Sunyaev}, R., {Arefiev}, V., {Babyshkin}, V., {et~al.} 2021, \aap, 656, A132

\bibitem[{{Sunyaev} \& {Zeldovich}(1972)}]{1972CoASP...4..173S}
{Sunyaev}, R.~A. \& {Zeldovich}, Y.~B. 1972, Comments on Astrophysics and Space
  Physics, 4, 173

\bibitem[{{Swetz} {et~al.}(2011){Swetz}, {Ade}, {Amiri}, {Appel},
  {Battistelli}, {Burger}, {Chervenak}, {Devlin}, {Dicker}, {Doriese},
  {D{\"u}nner}, {Essinger-Hileman}, {Fisher}, {Fowler}, {Halpern},
  {Hasselfield}, {Hilton}, {Hincks}, {Irwin}, {Jarosik}, {Kaul}, {Klein},
  {Lau}, {Limon}, {Marriage}, {Marsden}, {Martocci}, {Mauskopf}, {Moseley},
  {Netterfield}, {Niemack}, {Nolta}, {Page}, {Parker}, {Staggs}, {Stryzak},
  {Switzer}, {Thornton}, {Tucker}, {Wollack}, \& {Zhao}}]{2011ApJS..194...41S}
{Swetz}, D.~S., {Ade}, P.~A.~R., {Amiri}, M., {et~al.} 2011, \apjs, 194, 41

\bibitem[{{Tinker} {et~al.}(2008){Tinker}, {Kravtsov}, {Klypin}, {Abazajian},
  {Warren}, {Yepes}, {Gottl{\"o}ber}, \& {Holz}}]{2008ApJ...688..709T}
{Tinker}, J., {Kravtsov}, A.~V., {Klypin}, A., {et~al.} 2008, \apj, 688, 709

\bibitem[{{Umetsu}(2020)}]{2020A&ARv..28....7U}
{Umetsu}, K. 2020, \aapr, 28, 7

\bibitem[{{van Breukelen} {et~al.}(2006){van Breukelen}, {Clewley}, {Bonfield},
  {Rawlings}, {Jarvis}, {Barr}, {Foucaud}, {Almaini}, {Cirasuolo}, {Dalton},
  {Dunlop}, {Edge}, {Hirst}, {McLure}, {Page}, {Sekiguchi}, {Simpson}, {Smail},
  \& {Watson}}]{2006MNRAS.373L..26V}
{van Breukelen}, C., {Clewley}, L., {Bonfield}, D.~G., {et~al.} 2006, \mnras,
  373, L26

\bibitem[{{van der Burg} {et~al.}(2016){van der Burg}, {Aussel}, {Pratt},
  {Arnaud}, {Melin}, {Aghanim}, {Barrena}, {Dahle}, {Douspis}, {Ferragamo},
  {Fromenteau}, {Herbonnet}, {Hurier}, {Pointecouteau},
  {Rubi{\~n}o-Mart{\'\i}n}, \& {Streblyanska}}]{2016A&A...587A..23V}
{van der Burg}, R.~F.~J., {Aussel}, H., {Pratt}, G.~W., {et~al.} 2016, \aap,
  587, A23

\bibitem[{{Vikhlinin} {et~al.}(2009){Vikhlinin}, {Burenin}, {Ebeling},
  {Forman}, {Hornstrup}, {Jones}, {Kravtsov}, {Murray}, {Nagai}, {Quintana}, \&
  {Voevodkin}}]{2009ApJ...692.1033V}
{Vikhlinin}, A., {Burenin}, R.~A., {Ebeling}, H., {et~al.} 2009, \apj, 692,
  1033

\bibitem[{{Warren} {et~al.}(2006){Warren}, {Abazajian}, {Holz}, \&
  {Teodoro}}]{2006ApJ...646..881W}
{Warren}, M.~S., {Abazajian}, K., {Holz}, D.~E., \& {Teodoro}, L. 2006, \apj,
  646, 881

\end{thebibliography}

\appendix

\section{(Weak) evolution of the $F_{\rm X}$-$M_{500}$ relation with redshift}
\label{sec:z05}

\begin{figure}[h!]
    \centering
    \includegraphics[angle=0,clip=true,width=0.95\columnwidth]{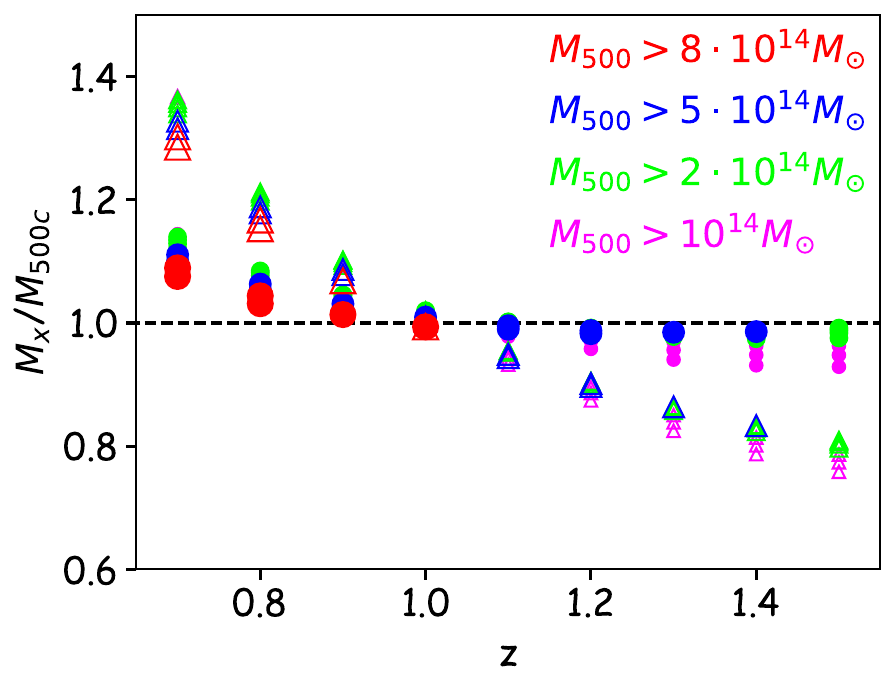}  
    \caption{Expected deviation of the X-ray-flux-based mass estimate from the true (input) mass. Triangles show the ratio of the mass predicted by eq.~(\ref{eq:CVS2015}) to the true $M_{500c}$ value  for clusters that follow adopted scaling relations from \cite{2009ApJ...692.1033V}. Solid circles show the same ratio when the factor $z^{0.5}$ (as in eq.~\ref{eq:xmass}) is accounted for. Different colors indicate the mass range of clusters that are expected to be present above a given redshift. }
    \label{fig:z05}
\end{figure}

We reproduce the analysis described in \cite{2015MNRAS.450.1984C}, namely, for each cluster mass $M_{\rm 500c}$ and redshift, we calculate the X-ray flux $F_{\rm X}$ using the scaling relations from \cite{2009ApJ...692.1033V} and assuming that they hold at redshifts of interest. We use the cluster mass function from \cite{2008ApJ...688..709T} to estimate the mass range relevant for a given redshift. Next, we convert the X-ray flux $F_{\rm X}$ to the cluster mass estimate $M_{\rm X}$ using Eq.~\ref{eq:CVS2015}. Figure~\ref{fig:z05} shows the ratio between $M_{\rm X}$ and the true cluster mass as a function of redshift. Only clusters with $M_{\rm 500c}>10^{14}$ $M_{\odot}$ are considered. As follows from Fig.~\ref{fig:z05}, there is a weak trend with redshift which can be accounted for to improve the tightness of the correlation between the X-ray flux and the cluster mass. 

\section{Biases of the simple aperture photometry}
\label{sec:bias}

Simple aperture photometry flux estimation can suffer from biases arising from fluctuations in the number of sources comprising the unresolved Cosmic X-ray Background (CXB). In this section, we briefly evaluate the magnitude of such bias using typical parameters relevant to the eROSITA all-sky survey and assess its importance.

For estimates, we used an effective exposure time of  $\sim 700\,{\rm s}$ and the time-averaged eROSITA background rates.  
The total telescope count rate is composed of three main components:  the instrumental background, the diffuse sky background (dominated by the Milky Way emission), and the CXB, which can be partially resolved in individual sources. On average, these three components have comparable contributions to the 0.4-2.3 keV band used in the previous sections. In Sec.~\ref{sec:Fx}, the background contribution to the cluster region ($2'$ radius circle) was estimated by measuring the surface brightness in a much larger region (a ring with inner and outer radii of $6'$ and $20'$, respectively) and multiplying it by the solid angle of the source region. While this procedure is expected to remove the particle and diffuse backgrounds accurately, the fluctuations associated with unresolved CXB sources introduce a bias in the measured flux. The bias arises due to a small number of CXB sources that can be detected and resolved with the eROSITA sensitivity and angular resolution. 

There are two flavors of this bias. The first type arises because resolving background sources in close vicinity to the target cluster (without compromising the cluster flux) is more difficult than farther away from it. For that reason, a higher flux threshold for excising CXB sources might be set in the `cluster' region compared to the `background'.  As a result, the mean surface brightness in the background region is lower than in the source region, even in the absence of a cluster, and the mean cluster flux is overestimated (after averaging over many clusters;  see Table~\ref{tab:bias}). However, the probability of finding a source with the flux in the relevant range in the cluster region is small. Therefore, this bias shifts the mean but does not change the mode of the distribution.  

The second type of bias stems purely from a much smaller solid angle of the `cluster' region than the `background' region. This means that the probability of finding a bright CXB source at a given flux level in the `cluster' region is smaller than in the `background' region. As a result, most of the `cluster' regions will lack such sources except for a few. Therefore, even if the CXB sources are removed at the same flux level and the mean surface brightness is the same, the most probable flux value (maximum of the probability distribution function) is lower in the smaller region. This is similar to the effects of low numbers statistics considered by \cite{2004MNRAS.351.1365G} in the context of calculating the total luminosity of a population of discrete sources in normal galaxies.

To illustrate the magnitude of these biases, we've made a simple simulation of the aperture photometry flux estimation. The particle and diffuse background levels (surface brightness) were set to typical values of the eROSITA survey. For CXB modeling, we used the source counts from the Chandra deep field \citep{2017ApJS..228....2L} in the 0.5-2~keV band. With these assumptions, $10^{6}$ realizations of the expected fluxes in the `cluster' and `background' regions were performed taking into account fluctuations in the number of CXB sources below the given detection threshold (and associated count rate in the eROSITA 0.4-2.3 keV band) to which the expectation of the particle and diffuse background rates were added. 

\begin{figure*}
    \centering
    \includegraphics[angle=0,clip=true,trim=0.5cm 5.cm 1cm 3cm,width=0.9\columnwidth]{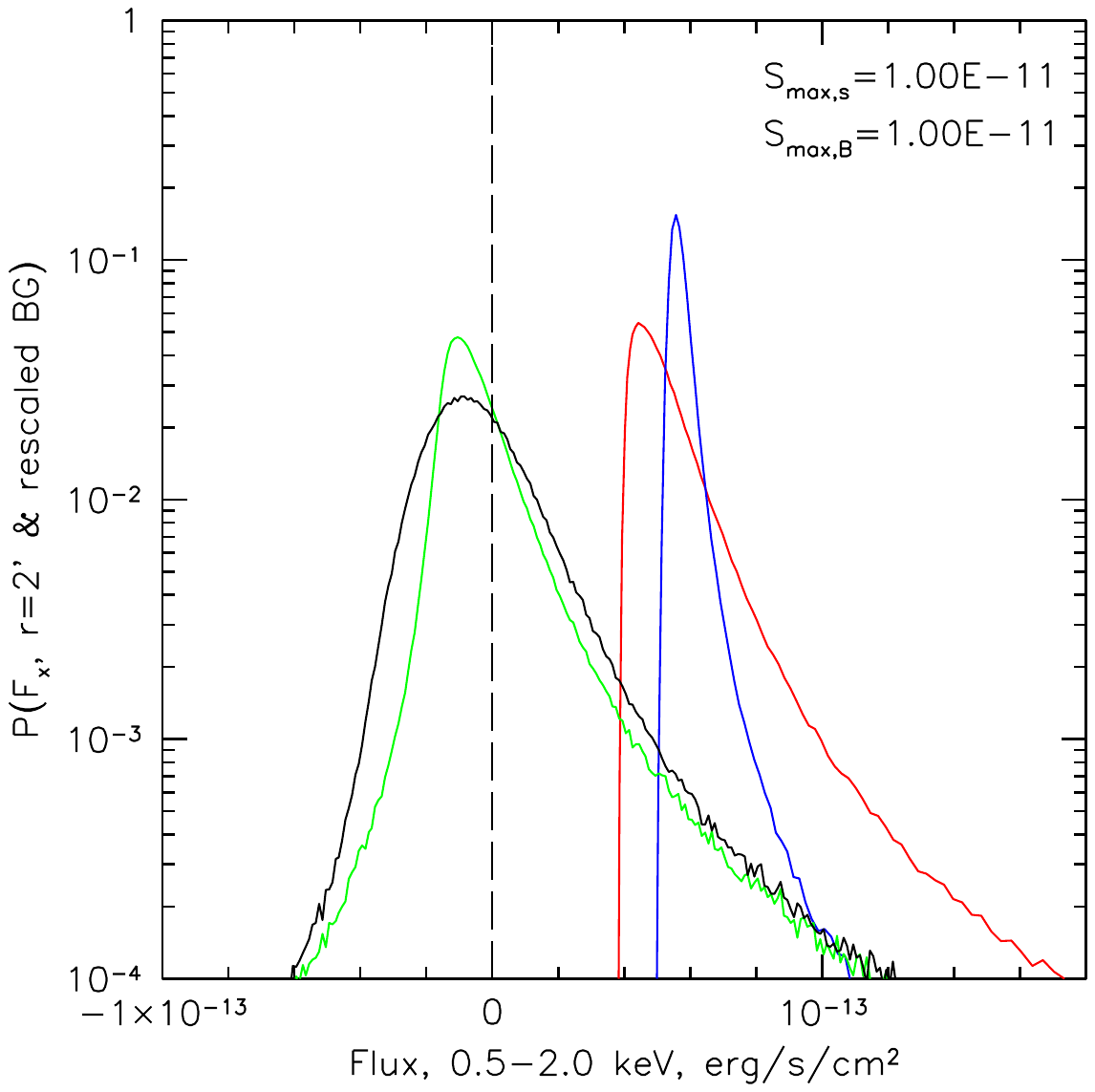}  \includegraphics[angle=0,clip=true,trim=0.5cm 5cm 1cm 3cm,width=0.9\columnwidth]{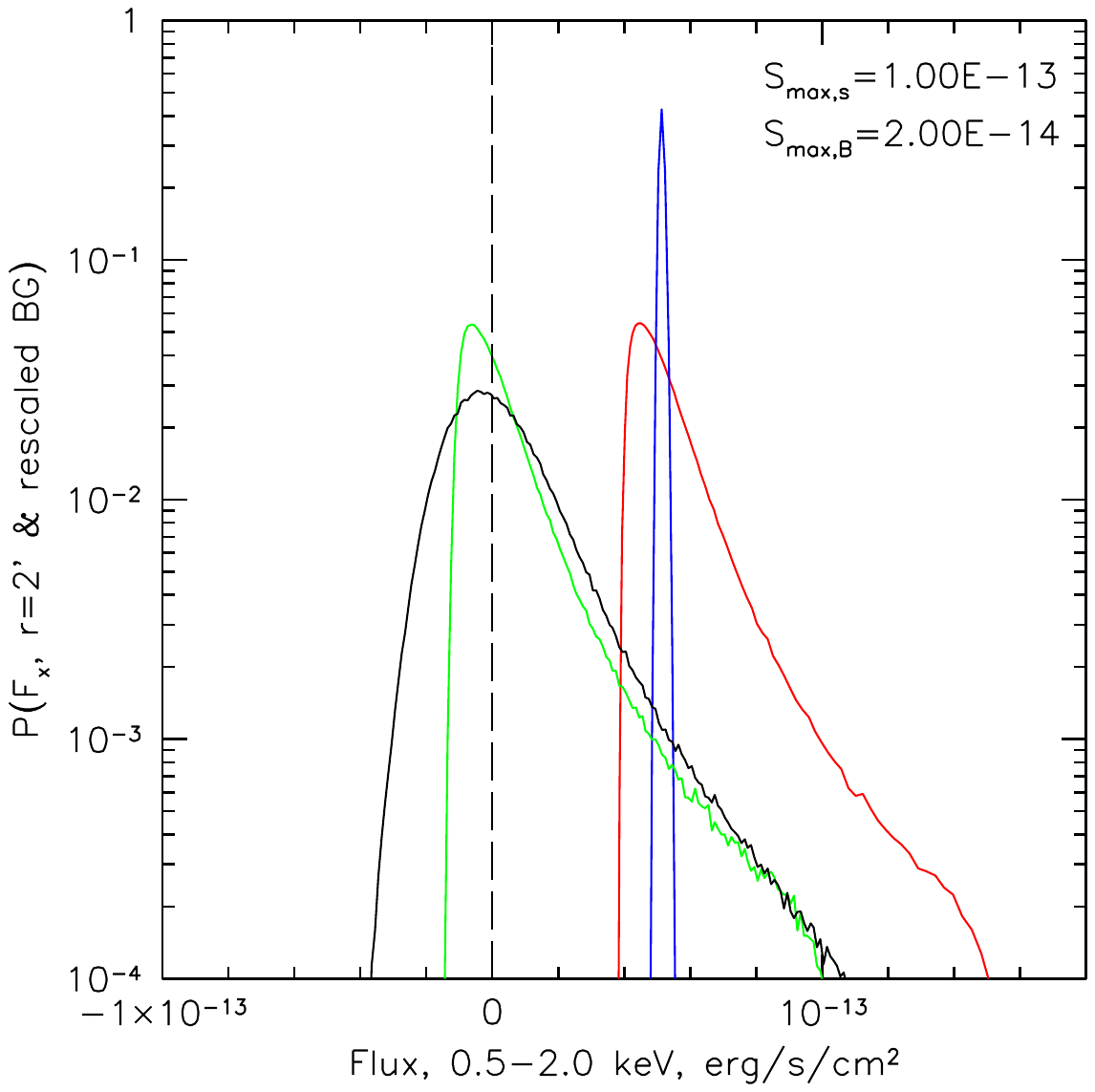}       
    \caption{Expected distribution of the 0.5-2.0~keV flux in a single circular `source' region ($2'$ radius) in the eROSITA all-sky survey (red curve) for different levels of the resolved sources in the source and background regions. The red curve corresponds to the distribution of fluxes (including diffuse and particle backgrounds) in the source region, while the blue curve shows the flux distribution in the background region ($\sim 100$ times larger solid angle) after rescaling to the area of the source region. These distributions are driven by fluctuations of CXB sources in these regions. The green curve shows the distribution of fluxes in the source region once the contribution of the background was subtracted. Finally, the black curve shows the same distribution to which photon counting noise has been added.  In the left panel, only extremely bright sources are excluded with $F_{\rm X}> 10^{-11} \,{\rm erg\,s^{-1}}$ in the $0.5-2\,{\rm keV}$ band. Such sources are very rare and do not contribute much to the mean CXB background. In practice, this flux cut effectively corresponds to the case when no CXB sources are resolved. In the right panel, fainter sources ($F_{\rm X}> 2\times 10^{-14} \,{\rm erg\,s^{-1}}$) have been removed from the background region, while in the source region, only sources brighter than $F_{\rm X}> 1\times 10^{-13} \,{\rm erg\,s^{-1}}$   have been removed. This reduces the scatter in the estimated flux in the background region but introduces a small bias in the mean flux since a larger fraction of the mean CXB flux is resolved in the background region.}
    \label{fig:bias}
\end{figure*}

The corresponding distributions are shown in Fig.\ref{fig:bias}  with the solid red and blue curves, while the green and black lines show the expected distributions of the flux corrected for the background without and with the account of the photon-counting-noise, respectively. Clearly, the expected distributions of fluxes (black curves) are broad, and their maxima are shifted below zero. In principle, the simulated distributions can be used to calculate the likelihood of measured fluxes in individual observations, taking into account different exposure times and different levels of the diffuse background. However, our aim is to estimate a single parameter - a normalization of the relation that links the measured X-ray flux with the mass. For this problem, the entire sample is used, and the biases can be estimated for the entire sample of 36 objects. This significantly increases the effective solid angle of the data and should reduce the scatter and some of the biases. 

\begin{table*}[h]
    \centering
    \caption{Estimated biases in fluxes measured from the $2'$ circles from simulations.}
    \begin{tabular}{l|r|r|r||r|r|l }
    $N_{\rm obj}$ &$F_{X,S}$ & $F_{X,B}$ & Mean & Mode (no PCN) & Mode (with PCN) \\
    \hline
      1 & $1\times 10^{-11}$ & $1\times 10^{-11}$ & $\sim 10^{-17}\sim 0$    &  $-1.1\times 10^{-14}$ &  $-8.5\times 10^{-15}$ & Fig.\ref{fig:bias} (left)\\
      1 & $1\times 10^{-11}$ & $2\times 10^{-14}$ & $+6.5\times 10^{-15}$    &  $-6.5\times 10^{-15}$ &  $-3.5\times 10^{-15}$ & \\
      1 & $1\times 10^{-13}$ & $2\times 10^{-14}$ & $+3.5\times 10^{-15}$    &  $-6.5\times 10^{-15}$ &  $-3.5\times 10^{-15}$ & Fig.\ref{fig:bias} (right)\\
      1 & $4\times 10^{-14}$ & $2\times 10^{-14}$ & $+1.8\times 10^{-15}$    &  $-6.5\times 10^{-15}$ &  $-3.5\times 10^{-15}$\\
      \hline
     36 & $1\times 10^{-11}$ & $1\times 10^{-11}$ & $\sim 10^{-17}\sim 0$    &  $-2.5\times 10^{-15}$ &  $-2.5\times 10^{-15}$ & \\
     36 & $1\times 10^{-11}$ & $2\times 10^{-14}$ & $+6.4\times 10^{-15}$    &  $+3.5\times 10^{-15}$ &  $+3.5\times 10^{-15}$ & \\
     36 & $1\times 10^{-13}$ & $2\times 10^{-14}$ & $+3.5\times 10^{-15}$    &  $+2.5\times 10^{-15}$ &  $+3.5\times 10^{-15}$ & Fig.\ref{fig:bias_sample}\\
     36 & $4\times 10^{-14}$ & $2\times 10^{-14}$ & $+1.8\times 10^{-15}$    &  $+1.5\times 10^{-15}$ &  $+1.5\times 10^{-15}$\\
     \hline
     1 & $1\times 10^{-11}$ & $1\times 10^{-13}$ & $ 2.8\times 10^{-15}$    &  $-9.5\times 10^{-15}$ &  $-6.5\times 10^{-15}$ & Fig.\ref{fig:bias_hilton}\\
     364 & $1\times 10^{-11}$ & $1\times 10^{-13}$ & $ 2.8\times 10^{-15}$    &  $1.5\times 10^{-15}$ &  $1.5\times 10^{-15}$ & Fig.\ref{fig:bias_hilton}\\
 \hline
    \end{tabular}
    \vspace{4pt}
    \tablefoot{All fluxes are in units of ${\rm erg\,s^{-1}cm^{-2}}$ in the 0.5-2~keV band. The bias in the mean flux (column `Mean') is caused by the different resolved fractions of CXB in the source and background regions, specified by the flux cuts $F_{\rm X,S}$ and $F_{\rm X,B}$, respectively. Another bias affects the mode of the measured flux distribution (columns `Mode (no PCN)'  and `Mode (with PCN)').  are caused by the fluctuations of the number of sources below the flux cut with and without account for a typical level of Photon Counting Noise (PCN) in eROSITA all-sky data. The rows of the Table marked with $N_{\rm obj}=1$ correspond to the biases in the distribution of fluxes for a single measurement, while those marked as $N_{\rm obj}=36$ characterize the mean flux in the sample of 36 objects. The last two rows correspond to the extended sample that features 364 objects, but sources with $F_{\rm X}>10^{-13}\rm {\rm erg\,s^{-1}cm^{-2}}$ are removed only from the `background region. In all cases, the distribution is calculated in flux bins of $10^{-15}\rm {\rm erg\,s^{-1}cm^{-2}}$ and, therefore, the mode has comparable uncertainty.}
    \label{tab:bias}
\end{table*}

\begin{figure*}
\begin{minipage}[t]{0.49\linewidth}
\centering
\includegraphics[angle=0, width=0.9\linewidth]{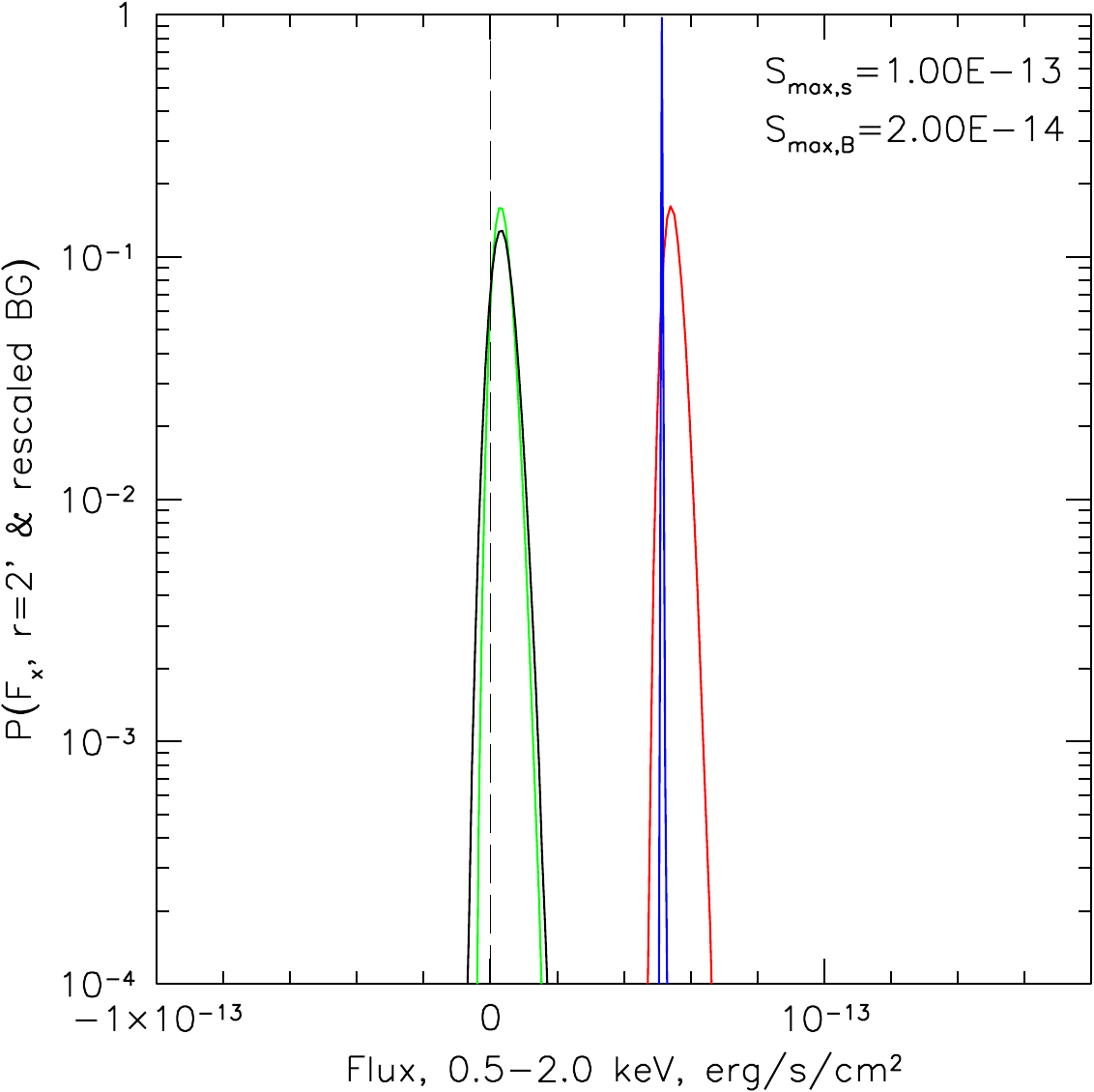}
\caption{Same as Fig.~\ref{fig:bias} (right panel) after averaging over a sample of $N_{\rm obj}=36$ objects. Now, the distributions are much narrower, and the mode and mean of the distributions agree well and feature only a small deviation from zero. This means that while individual flux measurements have a broad and biased distribution, the mean flux of the sample has a significantly smaller bias that is largely driven by the different resolved fractions of the CXB.}
\label{fig:bias_sample}
\end{minipage}
\hfill
\begin{minipage}[t]{0.49\linewidth}
\centering
\includegraphics[angle=0, width=0.9\linewidth]{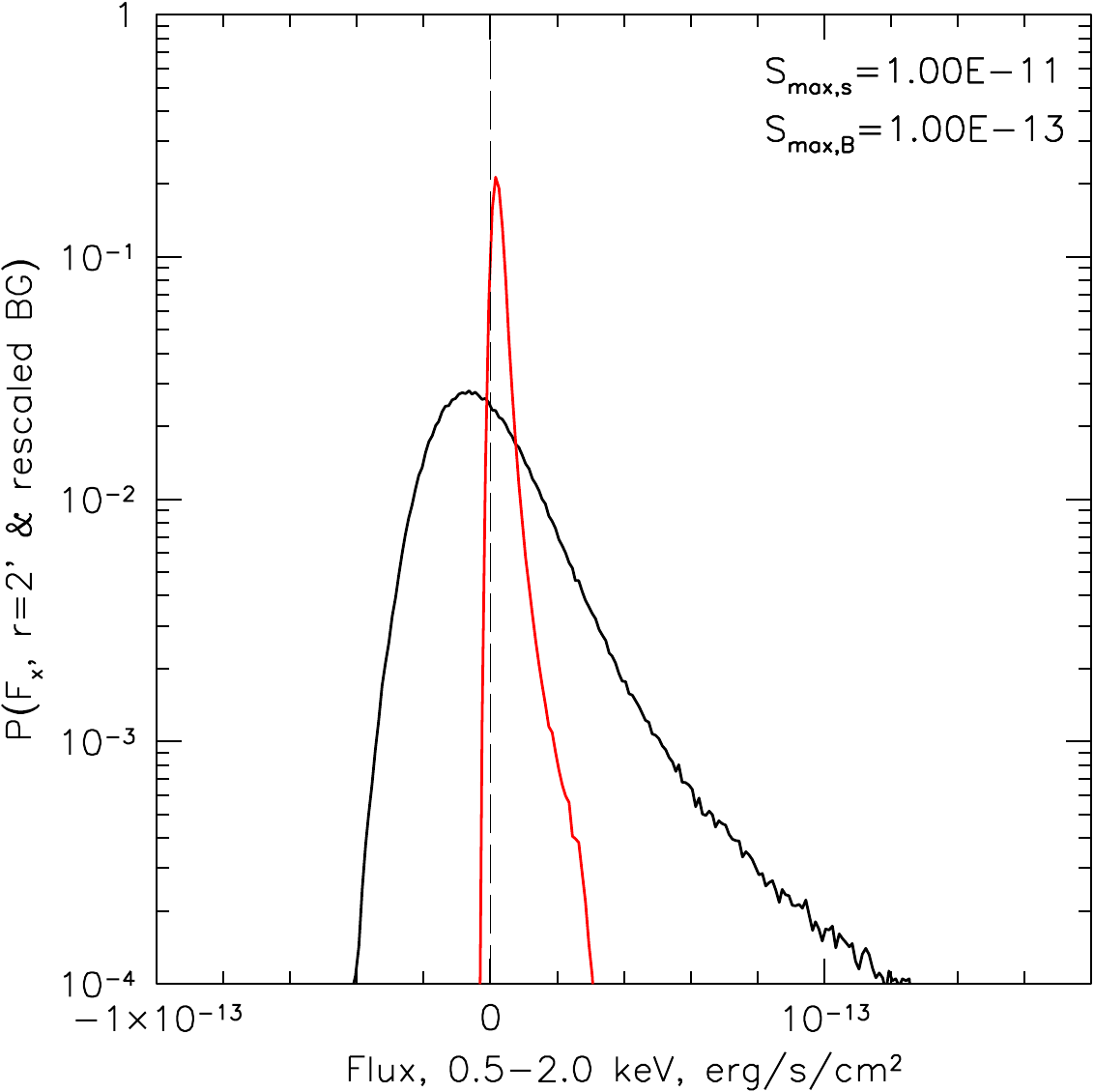}
\caption{Simulated distribution of the background-subtracted fluxes in the $2'$ circle apertures for a sample of 364 objects when CXB sources with fluxes above $10^{-13}\,{\rm erg\,s^{-1} cm^{-2}}$ are resolved in the background region. The black line corresponds to a single realization, while the red line applies to the sample-averaged flux.}
\label{fig:bias_hilton}
\end{minipage}%
\end{figure*}

To this end, we made additional simulations that account for the effective solid angle of the entire sample rescaled as shown in Fig.~\ref{fig:bias_sample}. In this case, the peak of the distribution becomes narrower and shifts to the positive side since the scatter associated with the fluctuations of unresolved sources is strongly reduced, but the bias associated with the different resolved fractions of the CXB background remains in place. Most importantly, the mean and the mode of the distribution now agree well.    Table~\ref{tab:bias} summarizes the biases discussed above. Our particular choice of the resolved fractions corresponds to cases shown in the right panel of Fig.\ref{fig:bias} (for a single object) and Fig.~\ref{fig:bias_sample} (for the entire sample). 

Similar simulations were done for larger sample ($N_{obj}=364$) with different CXB resolved fractions, namely $F_{\rm X,S}>10^{-11} \,{\rm erg\,s^{-1}}$ and $F_{\rm X,B}>10^{-13} \,{\rm erg\,s^{-1}}$ for the source and background regions, respectively. The mean and mode of the flux distribution for a single measurement and for the entire sample are shown in the last two rows of Table~\ref{tab:bias}, while the distributions themselves are shown in Fig.~\ref{fig:bias_hilton}. As expected, with these flux cuts, the flux distribution in a single measurement is broad, and its mode is shifted towards negative fluxes. In contrast, for the entire sample, the mode is much smaller and positive.   

The choice of X-ray flux estimation method (forced photometry) and the simplicity of the fitting function (only the normalization is treated as a free parameter) greatly simplifies the modeling of the uncertainties associated with the removal of the astrophysical background (CXB). As already mentioned in Sec.~\ref{sec:calib}, accurate constraints on the full scaling relation require full modeling (direct simulations) of the flux measurements, the intrinsic scatter in the cluster properties, and the biases in the selection function.

\label{lastpage}
\end{document}